\documentclass[twocolumn,prx,showpacs,amsmath,amstex,amssymb,mathfonts,superscriptaddress]{revtex4-1}

\usepackage{amsthm,color,amsfonts,graphicx,verbatim}
\usepackage{amsmath}
\usepackage{amssymb}
\usepackage{amsthm}
\usepackage{amsfonts}
\usepackage{verbatim}
\usepackage{enumerate}
\usepackage{latexsym}

\usepackage[normalem]{ulem}

\usepackage{bm}
\usepackage{graphicx}
\usepackage{dsfont}

\usepackage{hyperref}
\usepackage[normalem]{ulem}

\hypersetup{
    bookmarks=true,         
    unicode=false,          
    pdftoolbar=true,        
    pdfmenubar=true,        
    pdffitwindow=false,     
    pdfstartview={FitH},    
    pdftitle={Quasi-adiabatic dynamics in periodically driven systems},    
    pdfauthor={},     
    pdfsubject={},   
    pdfcreator={},   
    pdfproducer={}, 
    pdfkeywords={keyword1} {key2} {key3}, 
    pdfnewwindow=true,      
    colorlinks=true,       
    linkcolor=black,          
    citecolor=blue,        
    filecolor=magenta,      
    urlcolor=blue           
}

\newcommand{\be}{\begin{equation}}
\newcommand{\ee}{\end{equation}}
\newcommand{\bea}{\begin{eqnarray}}
\newcommand{\eea}{\end{eqnarray}}

\newcommand{\la}{\langle}
\newcommand{\ra}{\rangle}

\renewcommand{\phi}{\varphi}
\renewcommand{\epsilon}{\varepsilon}

\newcommand{\ww}{\textcolor{blue}}

\begin{document}
\title{Quasi-adiabatic dynamics and state preparation in Floquet many-body systems}

\author{Wen Wei Ho}
\affiliation{Department of Theoretical Physics, University of Geneva, Geneva, Switzerland}
\affiliation{Kavli Institute for Theoretical Physics, University of California, Santa Barbara, CA 93106, USA}

\author{Dmitry A. Abanin}
\affiliation{Department of Theoretical Physics, University of Geneva, Geneva, Switzerland}
\affiliation{Kavli Institute for Theoretical Physics, University of California, Santa Barbara, CA 93106, USA}

\date{\today}
\begin{abstract}

Periodic driving has emerged as a powerful experimental tool to engineer physical properties of isolated, synthetic quantum systems. However, due to the lack of energy conservation and heating effects, non-trivial (e.g., topological) many-body states in periodically driven (Floquet) systems are generally metastable. Therefore it is necessary to find strategies for preparing long-lived many-body states in Floquet systems. We develop a theoretical framework for describing the dynamical preparation of states in Floquet systems by a slow turn-on of the drive. We find that the dynamics of the system is well approximated by the initial state evolving under a slowly varying effective Hamiltonian $H_{\rm eff}^{(s)}(t)$, provided the ramp speed $s \gg t_*^{-1} \sim e^{-{\mathcal{C} \frac{\omega}{J}}}$, the inverse of the characteristic heating time-scale in the Floquet system. At such ramp speeds, the heating effects due to the drive are exponentially suppressed. We compute the slowly varying effective Hamiltonian $H_{\rm eff}^{(s)}(t)$, and show that at the end of the ramp it is identical to the effective Hamiltonian of the unramped Floquet system, up to small corrections of the order $O(s)$. 
As an application, we consider the passage of the slow quench through a quantum critical point (QCP), and estimate the energy absorbed due to the non-adiabatic passage through the QCP via a Kibble-Zurek mechanism. By minimizing the energy absorbed due to both the drive and the ramp, we find an optimal ramp speed $s_* \sim t_*^{-z/({d+2z})}$ for which both heating effects are exponentially suppressed. Our results bridge the gap between the numerous proposals to obtain interesting systems via Floquet engineering, and the actual preparation of such systems in their effective ground states.



\end{abstract}

\maketitle

\section{Introduction}


An important problem which arises across different fields of physics is to describe the behavior of quantum systems with parameters that vary periodically in time. In such systems, there exist so-called Floquet eigenstates which are stationary under stroboscopic time evolution (i.e. at times which are integer multiples of the driving period $T$), even though there are no truly stationary energy eigenstates like those in time-independent, static systems. Floquet eigenstates of a driven system can be very different from energy eigenstates of an undriven system; hence, periodic driving of a static system serves as a useful tool to realize phases of matter in different classes from that of the original system, some of which include new phases that are intrinsic to Floquet systems. 

Indeed, there has been much interest in Floquet engineering, both experimentally~\cite{Arimondo07,Obertaler08,Arimondo09,PhysRevLett.111.185301} and theoretically \cite{Eckardt05,PhysRevB.79.081406, PhysRevB.82.235114, Eckardt10, 2011NatPh...7..490L, Kollath11, Kitagawa11, PhysRevX.3.031005, PhysRevB.92.125107,Goldman14, 1367-2630-17-9-093039}. For instance, periodic driving has found a number of interesting applications in cold atoms experiments \cite{RevModPhys.80.885}. In particular, by using high frequency driving to control hopping parameters and to create artificial gauge fields, topological two-dimensional band structures in which Bloch bands had non-zero Chern numbers were recently realized in optical lattices \cite{2014Natur.515..237J, 2015NatPh..11..162A}. It has also further been proposed that high-frequency driving provides an avenue to realize intricate many-body states, such as fractional Floquet Chern insulators \cite{PhysRevLett.112.156801}. 

\begin{figure}[t]
\center
\includegraphics[width=0.5\textwidth]{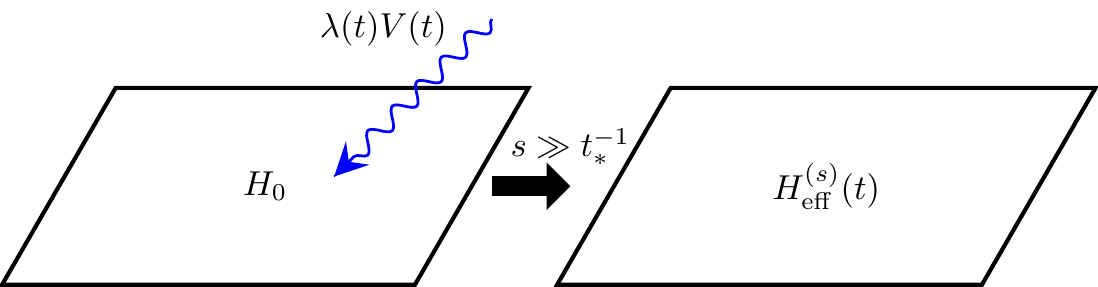}
\caption{We consider a set-up in which a static Hamiltonian $H_0$ (left) is subject to a high-frequency driving term $V(t) = V(t+T)$ with the amplitude $\lambda(t)$ slowly ramped from $0$ to $1$ over time $t = -\infty$ to $t = 0$. The speed of the ramp is characterized by parameter $s$, Eq.~(\ref{eq:condition_lambda}). Provided that $s \gg  t_*^{-1} \sim e^{-\mathcal{C} \frac{\omega}{J}} $, where $t_*$ is the characteristic heating time-scale, the system's stroboscopic evolution can be  accurately described by a slowly-varying effective Hamiltonian $H_{\rm eff}^{(s)}(t)$ (right).} 
\label{fig:system}
\end{figure}

While a promising tool to realize various phases of quantum matter, Floquet engineering in isolated many-body systems is hampered by an intrinsic difficulty: quantum phases of matter arise in ground states of physical Hamiltonians; however, the absence of energy conservation under periodic driving invalidates the very notion of a ground state. What state then, is one realistically targeting in Floquet engineering? It turns out that, at least for the case of high frequency driven systems, there exists a static effective Hamiltonian $H_{\rm eff}$ which describes approximate stroboscopic evolution for a parametrically long time in the driving frequency, the meaning and bounds of which have recently been established \cite{2015arXiv150905386A, 2015arXiv151003405A, PhysRevLett.116.120401, Kuwahara201696}. Hence, $H_{\rm eff}$ is a quasi-conserved quantity and its eigenstates are quasi-stationary states; the commonly-employed Floquet-Magnus expansion  \cite{Goldman14,Blanes2009151, 1367-2630-17-9-093039, 2015AdPhy..64..139B} can be used to generate it. Crucially, $H_{\rm eff}$ is a physical, local Hamiltonian, and so the targeted state in Floquet engineering is the ground state of this Hamiltonian. It should be noted that $H_{\rm eff}$ is in general quite different from the Floquet Hamiltonian $H_{F} \equiv \frac{i}{T} \log \mathcal{T} e^{-i \int_{0}^{T} dt H(t)}$ which describes exact stroboscopic time evolution. The latter is typically not a physical, local Hamiltonian and hence has no ground state; instead all its eigenstates look like infinite-temperature states~\cite{PhysRevE.90.012110,Ponte2015196, PhysRevX.4.041048}, reflecting the fact that generically (barring many-body localized systems~\cite{PhysRevLett.114.140401, LazaridesMBL, AbaninDrivenMBL}) all initial states eventually heat up indefinitely under the drive, although the time-scales involved can be very long~\cite{AbaninSlowHeating}.

Having identified the ground state  of $H_{\rm eff}$ as the one to target in Floquet engineering, how does one prepare the system in that state? It was proposed that one could perform a slow ramping up of the periodic drive starting from the ground state of the undriven system~\cite{Eckardt05,Kollath11,Kitagawa11}, and hope that the system `adiabatically' evolves towards the desired state. However, theoretical tools for analyzing the dynamics of the system under such a protocol are currently missing. 
 Even if there were some notion of `adiabatic' evolution, the unbounded heating under the drive suggests that we cannot turn on the ramp too slowly -- if so, the system will eventually absorb so much energy from the drive that it will heat up to infinite temperature, and no useful quasi-stationary state can be produced. It is hence important to theoretically understand the quasi-adiabatic dynamics that arises from a slow ramp of a Floquet system. Does this protocol work, and if so, what is the optimal ramp rate $s_*$ which allows one to get as close to the ground state of $H_{\rm eff}$ as possible?  In this paper, we develop a theoretical framework for addressing these questions. 

\subsection{Set-up}
We analyze the following typical situation, illustrated in Fig.~\ref{fig:system}: (i) At $t = -\infty$ the system is prepared in the ground state $| \psi_0 \ra$ of a static local many-body Hamiltonian $H_0$, which has a typical local energy scale of $J$. When periodically driven by a suitably chosen term $V(t) = V(t+T)$ at high frequencies $\omega = \frac{2\pi}{T} \gg J$, a quasi-conserved effective Hamiltonian $H_{\rm eff}$ is generated, whose ground state realizes a different phase   from that of $H_0$. Of particular interest is the case when the ground state of $H_{\rm eff}$ is topologically non-trivial. (ii) We ramp the strength $\lambda(t)$ of the drive, as is done in experiments, up to its maximal value at a final time $t = 0$. That is, the full time-dependent Hamiltonian is given by
\be\label{eq:hamiltonian}
H(t)=H_0+\lambda(t)V(t), \,\, \lambda(-\infty)=0, \,\, \lambda(0)=1,
\ee
and the system undergoes a unitary evolution with the Hamiltonian $H(t)$. (iii) We assume that the ramp is slow compared to the driving frequency:
\be\label{eq:condition_lambda}
\frac{d\lambda(t)}{dt}\sim s\ll \omega, \,\,\,  \forall t. 
\ee
We are interested in the time-evolved state $| \psi(t) \ra =  \mathcal{T} e^{-i \int_{-\infty}^t dt' H(t') } | \psi_0 \ra$, and in particular, the state $| \psi(0) \ra$ produced at the end of the ramp.

\subsection{Results}
Our main results are as follows. We find that there is an effective slowly varying Hamiltonian $H_{\rm eff}^{(s)}(t)$ which accurately describes stroboscopic dynamics throughout the ramp, provided that the ramp rate $s$ is much larger than the inverse of the heating time $t_*$ (which is exponentially long in the driving frequency~\cite{AbaninSlowHeating}),
\be
s \gg  t_*^{-1} \sim e^{-\mathcal{C} \frac{\omega}{J} }.
\label{eqn:heating}
\ee
Here $H_{\rm eff}^{(s)}(t)$ is slowly varying on the time scale $t\sim s^{-1}$, and is such that at the beginning of the ramp it is the original undriven Hamiltonian, $H_{\rm eff}^{(s)}(-\infty) = H_0$, while at the end of the ramp it is equivalent to the static effective Hamiltonian of the driven but unramped model, $H_{\rm eff}^{(s)}(0) = H_{\rm eff}$, up  to corrections of order $s/\omega $. 

Thus, even though there is no true `adiabatic limit' $s \to 0$, as long as $s \gg t_*^{-1}$, beginning from the ground state $| \psi_0 \rangle$ of $H_0$, the ramped Floquet system can be effectively understood as undergoing a slow quench from $H_0$ to $H_{\rm eff}$, for which an understanding of the dynamics exists. The resulting state $| \psi(0) \rangle$ will then be a good approximation to the ground state of $H_{\rm eff}$.

 As an application of this result, we consider the experimentally and theoretically interesting scenario where the ramp goes past a quantum critical point (QCP), illustrated in Fig.~\ref{fig:function}. Passing through a QCP is inevitable if one is attempting to realize a state $|\psi(0)\ra$  that is in a different universality class than that of the initial state $| \psi_0 \rangle$.  By minimizing the energy absorbed due to both the periodic drive and also the non-adiabatic passage through the QCP, we find an optimal ramp rate 
\begin{align}
s_*   \sim t_*^{-\frac{z}{d+2z}},
\end{align}
for which both heating effects (due to fast driving and slow ramp of $H_{\rm eff}^{(s)}(t)$) are exponentially small in frequency. Here $d$ is the dimensionality of the system and $z$ the critical exponent associated with the correlation length as one approaches the QCP. Thus, it is possible to prepare the `ground state' of a Floquet system with a high degree of control and accuracy.

Intuitively, our result can be understood as follows: there are two time-scales involved in the problem, a slow one set by the ramp speed $(t \sim s^{-1} )$ and a fast one set by the high-frequency driving $(t \sim \omega^{-1})$. We can `integrate' out the fast modes leading to an effective slowly varying description of the problem. However, this procedure cannot fully eliminate all the fast modes; the residual fast terms eventually heat the system up, which leads to the heating rate $t_*$ that is exponentially small in frequency,  Eqn.~(\ref{eqn:heating}). If the ramp speed is sufficiently fast, $s \gg  t_*^{-1}$, the system will not absorb too much energy from the drive and can be described by a slowly varying Hamiltonian, even though the total ramp time is infinite. Once the problem has been reduced to that of a slow quench, the scaling of the local heat absorbed during a ramp through a quantum critical point with the ramp speed follows from the Kibble-Zurek mechanism. The optimal ramp rate $s_*$ can then be obtained by minimizing the local heating due to both the drive and the ramp.

In what follows, we develop this theoretical approach, based on a series of local, time-dependent, unitary transformations, which allows us to decouple fast oscillatory motion with frequency $\omega$ from the slow motion due to ramping up the strength of the drive. Using this, we demonstrate that the evolution of the system is well-approximated by the evolution with a slowly varying Hamiltonian, which we calculate. Therefore, the system can stay close to the ground state of the effective Hamiltonian in a broad interval of ramp speeds.

 We emphasize that our approach, described below in detail, is conceptually very different from the conventional adiabatic Floquet theory (AFT), which is often employed for few-particle driven systems (for a recent review, see \cite{2016arXiv160602229W}; see also \cite{PhysRevLett.101.245302, 2016arXiv160608041E}). In AFT, one follows the adiabatic evolution of Floquet eigenstates: one prepares the system initially in some Floquet eigenstate and tracks whether the system remains in or near an instantaneous Floquet eigenstate. However, in our work, we initialize our system in the ground state of $H_0$ and show that the system tracks instead the ground state of the slowly varying effective Hamiltonian $H_{\rm eff}^{(s)}(t)$, rather than a Floquet eigenstate of $H_F$, which we stress are not equivalent. Moreover, AFT is inapplicable to experimental efforts to prepare interesting many-body states, since as discussed above, one is targeting the ground state of $H_{\rm eff}$ in Floquet engineering, and not an infinite-temperature Floquet eigenstate. We note that the authors of a recent paper~\cite{0295-5075-115-3-30006} claimed that there is a Floquet eigenstate of $H_F$ that is truly a ``ground state", i.e.~which shows ``many properties similar to the ground states of a static Hamiltonian", such as area-law entanglement entropy. We believe that this is incorrect, as it contradicts the general arguments of Refs.~\cite{PhysRevE.90.012110,Ponte2015196, PhysRevX.4.041048}, which show that {\it all} Floquet eigenstates must be effectively infinite temperature states.

\subsection{Structure of paper}

The paper is organized as follows. In Sec.~\ref{sec:iterative}, we describe the gauge transformation which allows us to iteratively integrate out the fast motion and to calculate the slowly varying effective Hamiltonian $H_{\rm eff}^{(s)}(t)$ in powers of $J/\omega$. We argue that this procedure is generically an asymptotic expansion, and derive the optimal order at which the procedure must be stopped to get the best results. We use this to estimate the local heating rate due to the remaining rapidly oscillating modes not eliminated and obtain the bound on the ramp rate $s \gg t_*^{-1}$. In Sec.~\ref{sec:FM} we compare $H_{\rm eff}^{(s)}(t)$ to $H_{\rm eff}$, the effective Hamiltonian of the unramped Floquet system, which can be calculated through a Floquet-Magnus expansion, and show they are identical, up to corrections of order $O(s/\omega)$. Then, in Sec.~\ref{sec:ramp}, we estimate the scaling of the energy absorbed through the Kibble-Zurek mechanism when $H_{\rm eff}^{(s)}(t)$ crosses a QCP, and derive the optimal ramp rate $s_*$ to minimize heating effects of the drive and ramp. Lastly, we in Sec.~\ref{sec:conclusion} we provide a summary of results and discuss some of their implications.


\section{Iterative process to generate time-varying effective Hamiltonian}
\label{sec:iterative}
Here we describe the iterative procedure to generate the time-varying effective Hamiltonian $H_{\rm eff}^{(s)}(t)$, and derive the range of validity of such a description.

\subsection{Mathematical preliminaries}
We first introduce some necessary mathematical assumptions and definitions. $H_0$ (and also $V(t)$) is assumed to be a many body operator with $R$-local interactions, that is, it is given by a sum of strictly local terms, $H_0 = \sum_X H_{0,X}$, where $X$ is the support of the local term $H_{0,X}$ whose linear dimension is bounded from above by $R$.  By the local norm or `size' of a many body operator $O(t) = \sum_X O_X(t)$, we mean $||O(t)||_\text{loc} \equiv \sup_{X,t} ||O_X(t)||$, where the norm on the right-hand side is the operator norm (a similar local norm was used in \cite{2015arXiv151003405A, 2015arXiv150905386A}). Thus, the typical local energy scale alluded to in the introduction is $J = \max \{ ||H_0||_\text{loc}, ||V(t)||_\text{loc}|| \}$. Since we will always work with the local norm, we will henceforth drop the subscript `loc'. 

It is also convenient to study the case when the ramp is exponential, 
$$\lambda(t)=e^{st}.$$ 
Such a choice makes formulas more compact, but our conclusions hold generally for other slow ramps, except for the results pertaining to the ramp through the QCP in Sec.~\ref{sec:ramp} which depend explicitly on the ramp protocol. 

We also note that the use of an infinite ramp time (from $t = -\infty$ to $0$) can be relaxed to that of a finite but large ramp time, i.e.~from $t = -t_i$ to $0$.  In this case, the results obtained still apply, except that now the effective dynamics of the system is described by an additional sudden quench from $H_0$ to $H_{\rm eff}^{(s)}(-t_i)$ initially at $t = t_i$, followed then by the slow quench dynamics which is the main result of the paper. The deviation of $H_0$ from $H_{\rm eff}^{(s)}(-t_i)$ is exponentially small in $t_i$, $||H_0-H_{\rm eff}^{(s)}(-t_i)|| \propto e^{-st_i}$, and can be made arbitrarily weak.


\subsection{Iterative process through a gauge transformation}
We proceed to deriving the slowly varying effective Hamiltonian $H_{\rm eff}^{(s)}(t)$ and describing the time evolved state $|\psi(t)\ra = \mathcal{T} e^{-i \int_{-\infty}^t dt' H(t') } |\psi_0\rangle$. As mentioned in the introduction, conceptually, our strategy is to separate motions on slow time-scales ($t\sim s^{-1}$) and fast time-scales ($t \sim \omega^{-1}$). To that end, we make a time-dependent unitary transformation which transforms the time-dependent Hamiltonian into a new Hamiltonian with a large part which varies slowly in time, and a small part which varies quickly. We will choose the transformation to minimize the norm of the rapidly oscillating part. A similar approach was taken in Refs.~\cite{Goldman14,2015arXiv151003405A, 2015arXiv150905386A}, but there it was used to derive the effective Hamiltonian of an unramped Floquet system.

Specifically, we perform a `gauge' transformation, acting on the wave function $| \psi(t) \ra$ in the lab frame with a unitary transformation $\hat Q^\dagger(t) \equiv e^{-\Omega(t)}$:
\be\label{eq:gauge}
\hat Q^\dagger(t)|\psi(t)\ra=|\phi(t)\ra,
\ee
to get a new wave function $| \phi (t) \rangle$ in a rotated frame. We choose $\Omega(-\infty)= \Omega(0)= 0$, so that $\hat Q(-\infty) = \hat Q(0) = \mathbb{I}$. This gives us the conditions that the initial and final states are identical in either frame of reference: $|\phi(-\infty)\ra=|\psi_0\ra$ and $|\phi(0)\ra=|\psi(0)\ra$. However, the wave function $| \phi(t) \ra$ in the rotated frame instead evolves under a transformed Hamiltonian:
\be\label{eq:hamiltonian2}
H'(t)=\hat Q^\dagger (t)  H(t)\hat Q(t)  - iQ^\dagger (t) \partial_t \hat Q(t),
\ee
and it is from a suitably transformed Hamiltonian that we can understand the quasi-adiabatic dynamics and results of the paper.

In the high-frequency limit, it is natural to expand $\Omega(t)$ as a power series in $J/\omega$, i.e.~$\Omega(t) = \sum_{k=1}^{n} \Omega^{(k)} (t)$, where  $\Omega^{(k)}(t) = -\Omega^{(k)}(t)^\dagger$ has norm $\sim (J/\omega)^k$. Here $n$  is some order at which the expansion is stopped. $\Omega^{(k)}(t)$ is perspicaciously chosen to remove the rapidly oscillating time-dependent terms of $H'(t)$ at order $k-1$ and leave behind slowly-varying time dependent terms at order $k$ and higher, and to also satisfy $\Omega^{(k)}(-\infty) = \Omega^{(k)}(0) = 0$. Thus, this process makes $H'(t)$ have a remaining rapidly oscillating time-dependent piece only at  $n$-th and higher orders. In a many-body system, the norm of this rapidly oscillating time-dependent term initially decreases due to the small prefactor $(J/\omega)^n$ acquired through the transformation, but we will see that it eventually increases due to the large number of terms generated as a result of the many nested commutators in its definition, reflecting that this procedure will produce an asymptotic series. There will therefore be some optimal order $n$ which we call $n_*$, where the norm of the remaining rapidly oscillating time-dependent term is smallest, which has to be calculated. The collection of the slowly-varying terms up to order $n_* - 1$ then represents the slowly-varying effective Hamiltonian $H_{\rm eff}^{(s)}(t)$. In what follows, we describe the systematic, iterative procedure that implements the desired transformations, and use this to derive $n_*$.
\newline

{\it Zeroth and first orders.} We first explicitly describe the procedure for the zeroth and first orders in $ J/\omega$, before stating the general form for higher orders. By going through the logic of the procedure for low orders, we will gain an understanding of the structure of the expansion.

We set $n = 2$ and the gauge transformation then takes the form $\hat{Q}(t) = e^{\sum_{k=1}^{2} \Omega^{(k)}(t) }$. Eqn.~(\ref{eq:hamiltonian2}) takes the following form:
\be\label{eq:hamiltonian3}
H'(t)=e^{-\Omega} H(t) e^{\Omega} -i \frac{1-e^{-{\rm ad}_\Omega}}{{\rm ad}_\Omega} \partial_t \Omega,
\ee
where ${\rm ad}_\Omega A=[\Omega, A]$. The zeroth order term in $J/\omega$, assuming $|| \Omega^{(1)} || \sim (J/\omega)$ is
\begin{align}
H_0 + H^{(0)}(t),
\label{eqn:zerothOrder}
\end{align}
where 
\begin{align}
& H^{(0)}(t)  \equiv G^{(0)}(t) - i \partial_t \Omega^{(1)}(t), \nonumber \\
& G^{(0)}(t) = e^{st} V(t).
\end{align}

Naively, one would get rid of the rapidly oscillating part by imposing $H^{(0)} = 0$; this means that
\begin{align}
\Omega^{(1)}(t) = -i \int_{-\infty}^{t} dt' e^{st'} V(t').
\end{align}
However, this would be inconsistent with the condition $\Omega^{(1)}(0) = 0$. Instead, we will allow for a slowly varying remaining piece, and so set
\begin{align}
H^{(0)}(t) = e^{st} W^{(0)},
\label{eqn:Wterm}
\end{align} 
where $W^{(0)}$ is a static term. Integrating Eqn.~(\ref{eqn:Wterm}) with the appropriate boundary condition defines both $W^{(0)}$ and $\Omega^{(1)}(t)$:
\begin{align}
& \Omega^{(1)}(t) = - i \int_{-\infty}^t dt' e^{st'}( V(t') - W^{(0)} ), \text{ and } \nonumber \\
& W^{(0)} = s \int_{-\infty}^0 dt' e^{s t'} V(t').
\label{eqn:Omega}
\end{align}

At this stage, having defined $W^{(0)}$ and $\Omega^{(1)}(t)$, we need to check that our assumptions of the relative sizes of the norms of the terms are consistent. That is, we need to check that $|| W^{(0)}|| \sim (J/\omega)^0$, and $|| \Omega^{(1)}(t) || \sim (J/\omega)^1$, up to  corrections which are $O( s/\omega)$. Since we will need to perform this consistency check this for all orders in the expansion, let us be more general, and check this for $W^{(k)}, \Omega^{(k)}(t)$. We claim that in these higher order terms, the relevant equations for $W^{(k)}, \Omega^{(k)}(t)$ have a similar form to Eqn.~(\ref{eqn:Omega}), substituting $V(t)$ with $V^{(k)}(t)$ and $e^{st}$ with $e^{pst}$.  Indeed, we are able to show (see Appendix A) that:
\begin{align}
| W^{(k)} ||  \leq &
\begin{cases}
||V^{(k)} ||\left(\frac{ps}{\omega} \right)  \qquad \text{ if } \bar{V}^{(k)} = 0 \\
||V^{(k)} ||\left(1 + \frac{ps}{\omega} \right) ~\text{ if } \bar{V}^{(k)} \neq 0
\end{cases},
\label{eqn:bound1} \\
|| \Omega^{(k)}(t) || &\leq 2 e^{p s t} || V^{(k)} || \left(\frac{1}{\omega} \right) \label{eqn:bound2},
\end{align}
where  $\bar{V}^{(k)} = \frac{1}{T}\int_0^T V^{(k)}(t) dt $ denotes the time average of $V^{(k)}(t)$. Hence, $|| W^{(k)} || \sim || V^{(k)} ||$ (if $\bar{V}{(k)} \neq 0$) and $||\Omega^{(k)}(t) || \sim \frac{1}{\omega} || V^{(k)} ||$. Importantly, we see from Eqn.~(\ref{eqn:bound2}) that  since $e^{p st} \leq 1$ throughout the ramp, the bound on $\Omega^{(k)}(t)$ is a uniform bound in $s$. 

Going back to the specific case of the zeroth order expansion we were considering, we see that $|| W^{(0)} || \sim J (J/\omega)^0 (s/\omega)$   because without loss of generality we can assume that the time average of the driving term is zero, $\bar{V} = 0$, and also $|| \Omega^{(1)}(t) || \sim (J/\omega)^1$. Therefore, the expansion is consistent. The magnitude of $W^{(0)}$ reflects that we have eliminated the rapidly oscillating terms but kept slowly varying terms, so it should be small only in $s/\omega$. Thus, the resulting effective Hamiltonian, with only slowly varying terms at zeroth order in $ J/\omega$, is given by Eqn.~(\ref{eqn:zerothOrder}),
\begin{align}
H_0 + e^{st} W^{(0)}.
\end{align}
We also note that $\Omega^{(1)}(t)$ can be decomposed as 
\begin{align}
& \Omega^{(1)}(t) = e^{st} \check{\Omega}^{(1)}(t), \nonumber \\
& \check{\Omega}^{(1)}(t) \equiv -i \int_{-\infty}^0 dt' e^{st'} \left( V (t'+t) - W^{(0)}  \right),
\label{eqn:OmegaDecomposition}
\end{align}
where $\check{\Omega}^{(1)}(t)$ is a periodic function of $t$, as can be checked easily (i.e.~let $t \to t+T$). In fact,
\be
\Omega^{(1)}(t) = 0 \text{ for } t \in T \mathbb{Z},
\ee
implying that $\hat{Q}(t) = \mathbb{I}$ for stroboscopic times, at least up to this order.

Let us next proceed to the first order in $J/\omega$. This term in $H'(t)$ is given by $H^{(1)}(t)$, where 
\begin{align}
 H^{(1)}(t)  & \equiv G^{(1)}(t) - i \partial_t \Omega^{(2)}(t), \nonumber \\
 G^{(1)}(t)  & = - e^{st} [\check{\Omega}^{(1)}(t) , H_0]  +  \nonumber \\
& e^{2st} \left( -[\check{\Omega}^{(1)}(t), V(t)] +   \frac{i}{2}  [ \check{\Omega}^{(1)}(t), \partial_t \check{\Omega}^{(1)}(t)] \right) \nonumber \\
& \equiv \sum_{p=1}^2 e^{p st} V^{(1)}_p(t).
\end{align}
The periodic terms $V^{(1)}_p(t)$ are defined to be those that are multiplied by the slowly varying function $e^{p st }$.

Like before, to get rid of the rapidly oscillating terms but to keep the slowly varying terms, we set and choose
\begin{align}
& H^{(1)}(t) = \sum_{p=1}^{2} e^{p st} W^{(1)}_p, \nonumber \\
& \Omega^{(2)}(t)  = - i \sum_{p=1}^2 \int_{-\infty}^t dt' e^{pst'}( V^{(1)}_p(t') - W^{(1)}_p ), \nonumber \\
&  W^{(1)}_p = p s \int_{-\infty}^0 dt' e^{p st'} V^{(1)}_p(t'),
\end{align}
for $p = 1,2$.

Thus, up to first order in $J/\omega$, the slowly varying part of the rotated Hamitonian $H'(t)$ reads
\begin{align}
H_0 + e^{st} W^{(0)} + \sum_{p=1}^2 e^{p st} W^{(1)}_p.
\end{align}
Similar to Eqn.~(\ref{eqn:OmegaDecomposition}), $\Omega^{(2)}(t)$ has the decomposition $\Omega^{(2)}(t) = \sum_{p=1}^2 e^{p st} \check{\Omega}^{(2)}_p(t)$, where $\check{\Omega}^{(2)}_p(t)$ are periodic functions of $t$, and $\Omega^{(2)}_p(t) = 0 $ for stroboscopic times.
\newline

{\it Higher orders.} At this point, a pattern has emerged and we can state the general formula, for arbitrary $n$. The full Hamiltonian $H'(t)$, obtained with the gauge transformation $\hat{Q}(t) = e^{\sum_{k=1}^n \Omega^{(k)}(t)}$, is given by
\begin{align}
H'(t) = H_0 + \sum_{k=1}^\infty H^{(k)}(t),
\end{align}
where $|| H^{(k)} || \sim (J/\omega)^k$,
\begin{align}
& H^{(k)} \equiv G^{(k)}(t) - i \partial_t \Omega^{(k+1)}(t),
\label{eqn:Hk}
\end{align}
with $G^{(k)}(t)$ expressed in terms of $\Omega^{(1)}(t), \cdots, \Omega^{(k)}(t)$:
\begin{align}
&G^{(k)}(t)= \sum_{q=1}^k \frac{(-1)^k}{k!} \sum_{\stackrel{1 \leq i_1, \cdots, i_q < k}{ i_1 + \cdots + i_q = k}} {\rm ad}_{\Omega^{(i_1)}} \cdots {\rm ad}_{\Omega^{(i_q)}} H(t) ~ +  \nonumber \\
&~ i \sum_{m=1}^{k} \sum_{q = 1}^{k+1 -m} \frac{(-1)^{q+1}}{(q+1)!} \sum_{\stackrel{1 \leq i_1,\cdots,i_q \leq k+1-m}{i_1 + \cdots + i_q = k+1 - m}}  \!\!\!\!\!\!\!\!\!\!\!\! {\rm ad}_{\Omega^{(i_1)}} \cdots {\rm ad}_{\Omega^{(i_q)}} \partial_t \Omega^{(m)} \nonumber \\
& \equiv \sum_{p=1}^{k+1} e^{p s t } V^{(k)}_p (t),
\label{eqn:G}
\end{align}
where $\Omega^{(k)}(t) = 0$ for $k > n$. The decomposition of $G^{(k)}(t)$ in terms slowly varying functions $e^{pst}$ multiplied by periodic functions $V^{(k)}_p(t)$, where $p = 1, \cdots, k+1$, follows readily from induction.

To remove the rapidly varying terms while keeping only slowly varying terms, we set
\begin{align}
H^{(k)}(t) = \sum_{p=1}^{k+1} e^{p st } W_p^{(k)},
\end{align}
and define $\Omega^{(k+1)}(t)$ and $W^{(k)}_p$ through
\begin{align}
& \Omega^{(k+1)}(t)  = - i \sum_{p=1}^{k+1} \int_{-\infty}^t dt' e^{p st'}( V^{(k)}_p(t') - W^{(k)}_p ), \nonumber \\
& W^{(k)}_p = p s \int_{-\infty}^0 dt' e^{p st'} V^{(k)}_p(t').
\label{eqn:Omega}
\end{align}
From the above, we then have
\begin{align}
\Omega^{(k+1)}(t) = 0 \text{ for } t \in T \mathbb{Z},
\end{align}
as can be checked easily,  generalizing Eqn.~(\ref{eqn:OmegaDecomposition}), and so $\hat{Q}(t) = \mathbb{I}$ for those times.
\newline

\subsection{Effective Hamiltonian, optimal order $n_*$ and range of validity}
The upshot of the technical procedure described in the previous section is the following. Since $\hat{Q}(t)$ is equal to the identity $\mathbb{I}$ at $t = T \mathbb{Z}$, $|\psi(t) \rangle =  |\phi(t)\rangle$ at those times, and so the stroboscopic motion of the system is exactly captured by the rotated Hamiltonian $H'(t)$, obtained by carrying out the gauge transformation described above up to order $n$. It is given by
\begin{align}
H'(t) & = H_0 + \sum_{k=1}^{n-1} \sum_{p=1}^{k} e^{pst} W^{(k)}_p + \sum_{k \geq n} H^{(k)}(t) \nonumber \\
& \equiv H_0 + H_{\rm slow}(t) + H_{\rm fast}(t).
\label{eqn:finalH}
\end{align}
As the name suggests, $H_{\rm slow}(t) \equiv  \sum_{k=1}^{n-1} \sum_{p=1}^{k} e^{pst} W^{(k)}_p$ is the slowly varying local Hamiltonian, having had all the rapidly oscillating terms removed, while $H_{\rm fast}(t) \equiv  \sum_{k \geq n} H^{(k)}(t)$ is the rapidly oscillating piece, which is comprised of terms that appear at order $n$ and higher.

Since there is an infinite number of terms $H^{(k)}(t)$ in $H_{\rm fast}(t)$ (for $k \geq n$), and each $H^{(k)}(t)$  is comprised of $k$ nested commutators of $\Omega^{(1)}(t), \cdots, \Omega^{(k)}(t), H(t)$, $H_{\rm fast}$ is a sum of local terms with larger and larger support. However, it can be shown that norm of each term decays exponentially with the size of its support; thus $H_{\rm fast}(t)$ is quasi-local. The proof of this is essentially identical to that in  Refs.~\cite{2015arXiv151003405A, 2015arXiv150905386A}, using the definition of $H^{(k)}(t)$ in Eqns.~(\ref{eqn:Hk}, \ref{eqn:G}) and also the important result that the bound on $\Omega^{(k)}(t)$ is uniform in $s$ (Eqn.~(\ref{eqn:bound2})) in order for the techniques used in the references to carry over; we thus only state the result here.

We now choose the optimal $n$ which we call $n_*$ such that the norm of $H_{\rm fast}(t)$ is approximately minimized, which will have physical consequences such as heating timescales in the system. Following Refs.~\cite{2015arXiv151003405A, 2015arXiv150905386A}, for $k \leq n-1$, we have
\begin{align}
|| \Omega^{(k+1)}|| \leq 4 \pi (2\pi C_0 R)^k k! J \left(\frac{J}{\omega}\right)^{k+1},
\label{eqn:OmegaNorm}
\end{align}
where $C_0$ is a combinatorial constant of order $1$, and $R$ is the range of the initially local Hamiltonian. We see that there are the two prefactors that lead to a competition of the growth of the norm of $ \Omega^{(k+1)}$: for small $k$ it decreases due to the small factor $(J/\omega)^{k+1}$ picked up during the iterative process, but for large $k$ there is a prefactor $k!$ that eventually dominates and the norm grows without bound. This prefactor $k!$ can be understood as arising due to the many-body nature of the problem: from its definition, Eq.~(\ref{eqn:Omega}) \& (\ref{eqn:G}), $\Omega^{(k+1)}$ involves $k$-nested commutators of many-body operators $H, V(t)$. The behavior of the growth of the norm of $\Omega^{(k+1)}$  reflects the fact that the gauge transformation $\hat{Q}(t)$ generically produces an asymptotic expansion.

 The optimal $n_*$ that approximately minimizes the norm of $H_{\rm fast}(t)$ is close to the one that minimizes the norm of $||\Omega^{(k+1)}||$ (see Refs.~\cite{2015arXiv151003405A, 2015arXiv150905386A} for the precise statement). From Eq.~(\ref{eqn:OmegaNorm}), we then have
\begin{align}
n_* \approx \frac{e^{-r}}{2\pi C_0 R J } \omega,
\label{eqn:OptimalN}
\end{align}
where $r = r(R) > 0 $ is a constant independent of $\omega$ (defined and used in \cite{2015arXiv151003405A, 2015arXiv150905386A}). At this order $n_*$, 
\begin{align}
||H_{\rm fast}(t)|| \leq Ce^{-r n_*},
\end{align}
with $C$ another constant depending on the microscopic details of the Hamiltonian such as $J$ and $R$ but also independent of $\omega$. Since $n_* \sim \omega$, we see that $||H_{\rm fast}(t)||$ is exponentially small in frequency.


By construction, the stroboscopic dynamics is exactly captured by $H'(t)$,  Eq.(\ref{eqn:finalH}), produced through the iterative procedure carried to this optimal order $n_*$. The smallness of the remaining rapidly oscillating term $H_{\rm fast}(t)$ indicates that energy absorption due to the drive is exponentially small, and so we instead only need to consider stroboscopic time evolution generated by the effective, slowly varying $n_*(R-1)$-local Hamiltonian,
\begin{align}
H_{\rm eff}^{(s)}(t) \equiv H_0 + H_{\rm slow}(t),
\label{eqn:effH}
\end{align} 
so that 
\begin{align}
|\psi(t) \ra \approx \mathcal{T} e^{-i \int_{-\infty}^t dt' H_{\rm eff}^{(s)}(t')} |\psi_0 \ra.
\label{eqn:approxPsi}
\end{align}

In order for this statement to be correct, we demand that the difference of the measurements of any local observable $O$ by a state evolved by $H'(t)$ and the same state evolved by $H_{\rm eff}^{(s)}(t)$ is small at the end of the ramp. That is, we require 
\be
\delta \langle O \rangle \equiv |\text{Tr}(U(0) \rho U^\dagger(0) O) - \text{Tr}(U_{\rm eff}^{(s)} (0) \rho U_{\rm eff}^{(s) \dagger}(0) O)| \ll 1.
\ee
Here $U(t)$ is the unitary time evolution operator generated by $H'(t)$ and $U_{\rm eff}^{(s)}(t)$ is the evolution operator generated by $H_{\rm eff}^{(s)}(t)$ (recall that both operators are equal to the identity at $t = -\infty$, not $0$).

We can bound the difference as (see Appendix B)
\begin{align}
& \delta \langle O  \rangle  \leq\int_{-\infty}^0 dt' \left| \left| [\mathcal{U}^{(s)}_{\rm eff}(t') O \mathcal{U}^{(s) \dagger}_{\rm eff}(t'), H_{\rm fast}(t')] \right| \right| ,
\end{align}
where $\mathcal{U}^{(s)}_{\rm eff}(t) = U_{\rm eff}^{(s) }(t) U_{\rm eff}^{(s) \dagger}(0)$, which is a time evolution operator that evoles a state from time $0$ to $t$ under $H_{\rm eff}^{(s)}(t)$. By using the Lieb-Robinson velocity \cite{Lieb1972, 2010arXiv1008.5137H} of $H_{\rm eff}^{(s)}(t)$ which goes as $\sim J$ to bound the physical growth of the operator $\mathcal{U}^{(s)}_{\rm eff}(t') O \mathcal{U}^{(s) \dagger}_{\rm eff}(t')$, and the fact that the norm of $H_{\rm fast}(t')$ is exponentially small, we can control this difference as 
\begin{align}
\delta \langle O\rangle  \leq  \sim e^{-r n_*(\omega)}  s^{-2},
\label{eqn:Odiff}
\end{align}
where we have kept only the scaling behavior of this quantity with $s$ and $\omega$. Desiring that the difference in the measurements is small, $\delta \la O \ra \ll 1$, gives us a condition on the ramp rate $s$:
\begin{align}
s \gg t_*^{-1} \sim  e^{-\mathcal{C} \frac{\omega}{J}},
\label{eqn:condition_s}
\end{align}
with $\mathcal{C}$ a $\omega$-independent constant, where we have used the fact that $n_*(\omega) \sim \omega$. We have also identified $t_* \sim e^{\mathcal{C} \frac{\omega}{J} }$ as the heating time scale appearing in Eq.~(\ref{eqn:heating}). 

Thus, the description of the system evolving under the slowly varying Hamiltonian $H_{\rm eff}^{(s)}(t)$ is valid throughout the ramp, as long as the ramp rate $s \gg  e^{-\mathcal{C} \frac{\omega}{J}}$. When $s <  e^{-\mathcal{C} \frac{\omega}{J}} $, we expect the system to heat up significantly during the ramp and the effective Hamiltonian description to break down; this  prevents an `adiabatic limit' $s \to 0$ in a Floquet system. From a physical standpoint, this result can be understood from the fact that heating rate in a Floquet system is exponentially small, so if the ramp time-scale $t \sim s^{-1}$ is much smaller than $t_*$, then the system would not have absorbed so much energy from drive and the effective Hamiltonian description is accurate.

\section{Connection to Floquet-Magnus expansion}
\label{sec:FM}

 At this stage, we have shown that the system can be accurately described by the slowly varying Hamiltonian $H_{\rm eff}^{(s)}(t)$ (Eqs.~(\ref{eqn:effH}), (\ref{eqn:approxPsi})) throughout the ramp provided $s \gg t_*^{-1}$, which can be generated through the iterative process described before. The optimal order $n_*$ to carry out this procedure to, is the same as that to produce the effective Hamiltonian for the unramped Floquet system, and only depends on the locality and typical energy scales of the original undriven Hamiltonian. Equivalently, the effective Hamiltonian of the Floquet system can also be obtained from the Floquet-Magnus expansion  \cite{Blanes2009151, 1367-2630-17-9-093039, 2015AdPhy..64..139B} up to the same order $n_*$.

However, we have yet to show that the effective Hamiltonian of the ramped Floquet system is related to that of the unramped Floquet system, whose ground state we would like to achieve. Intuitively, it is clear that this must be the case, and that the two effective Hamiltonians only differ by small terms proportional to $\sim s/\omega$, since these are the terms coming from the slowly varying part of the Hamiltonian after the rapidly oscillating terms have been integrated out. 

To show that this intuition is indeed correct, we now compute and compare both effective Hamiltonians of the ramped and unramped systems. We compute the Hamiltonians for the zeroth and first order in $(J/\omega)$, which will be sufficient for us to understand the general result (also, most Floquet engineering proposals only utilize the first order corrections due to the drive):
\begin{align}
H_{\rm eff}^{(s)}(t) &  = H^{(s,0)}(t) + H^{(s,1)} + \cdots, \nonumber \\
H_{\rm eff} & = H^{(0)} + H^{(1)} + \cdots,
\label{eqn:Expansion}
\end{align}
where the ramped Hamiltonian is parameterized by $s$, the ramp speed.

We switch to the Fourier representation of the problem, and denote the Fourier harmonics of $V(t)$ by $V_n$, $n\neq 0$ (without loss of generality the $n=0$ component can be always absorbed into $H_0$):
\be\label{eq:fourier}
V(t)=\sum_{n\neq 0} V_n e^{i\omega_n t}, \,\, \omega_n=\omega n, \,\,\, V_n^\dagger=V_{-n}.
\ee
We follow the prescription given in the previous section and only show the results. The zeroth order terms are
\begin{align}
H^{(s,0)}(t) &= H_0 + s e^{st}  \sum_{n \neq 0} \frac{V_n}{i\omega_n + s}, \nonumber \\
\label{eqn:H0}
H^{(0)} & = H_0,
\end{align}
\begin{widetext}
while the first order terms  are
\begin{align}
H^{(s,1)}(t)  & = -i e^{st}  \sum_{n \neq 0} \frac{[V_n,H]}{i \omega_n + s}  \left( 1 -  \frac{s}{( i\omega_n + s )} \right) +  \frac{i}{2}e^{2 st} \sum_{n \neq 0} \frac{[V_n,V_{-n}]}{ i \omega_n + s}  \nonumber \\
& + \frac{i}{2} e^{2st}\left ( \sum_{\stackrel{n, m\neq -n}{n  \neq 0, m \neq 0}} \frac{[V_n,V_{m}]}{(i \omega_n + s)} \frac{2s}{(i \omega_{n+m} + 2s)} - \sum_{n\neq 0, m \neq 0} \frac{[V_n,V_m]}{(i\omega_n + s)} \frac{2s}{(i \omega_m + 2s)} \left(1 + \frac{s}{(i \omega_m + s)} \right) \right),  \nonumber \\
H^{(1)} & = -i \sum_{n \neq 0} \frac{[V_n,H]}{i \omega_n} + \frac{i}{2} \sum_{n \neq 0} \frac{[V_n,V_{-n}]}{i \omega_n}.
\label{eqn:firstOrder}
\end{align}
\end{widetext}

From these expressions, it is evident that there is a lot of structural similarity between $H^{(s,k)}(0)$ and $H^{(k)}$. Notably, we see that $H^{(k)}$, with the renormalized energy denominators $i \omega_{\sum_{i=1}^p n_i } \to i \omega_{\sum_{i=1}^p n_i } + p s $, is contained within $H^{(s,k)}(0)$. We also see that the norm of the difference $|| H^{(s,k)}(0) -  H^{(k)}|| \sim s/\omega $. Even though we have only showed this to low orders, it is straightforward to see that this is true for higher orders too. 

This result is easy to understand: the system is not strictly periodic with the introduction of a ramp, and thus there will be perturbative corrections on top of the effective Hamiltonian $H_{\rm eff}$ of the Floquet system which scale as $\sim s/\omega$.  In the limit when the ramp speed $s \to 0$, $H^{(s)}_{\rm eff}(0) \to H_{\rm eff}$, which is expected. However, as discussed above, the effective Hamiltonian description will break down if $s$ is taken too small, due to heating effects.

Let us recapitulate. We have demonstrated that for ramp rates $s \gg e^{-\mathcal{C} \frac{\omega}{J} }$, the system can be effectively understood as undergoing a slow $(t \sim s^{-1}$) quench from $H_0$, the undriven Hamiltonian, to the effective Hamiltonian of the Floquet system, $H_{\rm eff}$ at time $t=0$. Therefore a ramped Floquet system also has `quasi-adiabatic' dynamics if we restrict ourselves to stroboscopic times, and we can borrow results from the large body of understanding of the adiabatic dynamics that arises for Hamiltonian systems undergoing a slow quench.

\section{Ramp through a quantum critical point and optimal ramp rate}
\label{sec:ramp}

\begin{figure}[t]
\center
\includegraphics[width=0.5\textwidth]{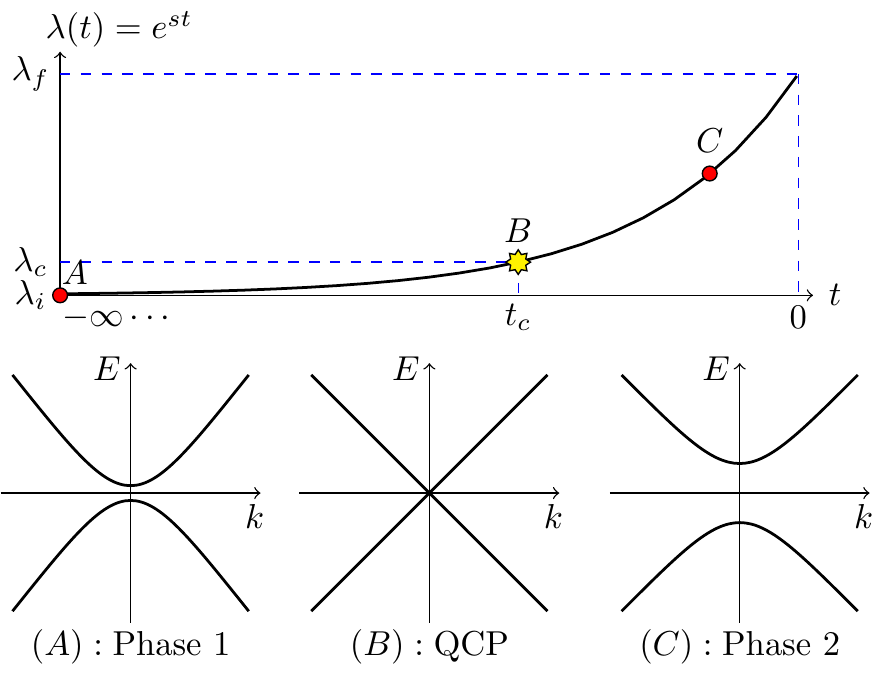}
\caption{(Top panel) The function considered in the slow ramp is $\lambda(t) = e^{st}$ which modulates the amplitude of the driving term $V(t)$. It starts from $\lambda_i = 0$ at $t = -\infty$ and ends at $\lambda_f = 1$ at $t = 0$. We also consider the scenario where the ramp goes through a quantum critical point (QCP), when the driving amplitude reaches a critical value $\lambda_c$ at time $t_c$, indicated on the diagram by point (B).  (Bottom panel) Schematic energy dispersion relations at points (A), (B), (C). At points (A) and (C), the system is gapped, describing phases $1$ and $2$ respectively, while at point (B), the system is at the QCP and is gapless. The points (A) and (B) can be arbitrarily close.} 
\label{fig:function}
\end{figure}

As an application of the previous result, let us now consider the case where the slowly varying effective Hamiltonian $H_{\rm eff}^{(s)}(t)$ crosses a quantum critical point (QCP), specifically a second order phase transition, at some critical time $t_c$. This situation is relevant both theoretically and experimentally: if we hope to use Floquet engineering to produce a state $| \psi(0)\ra$ that is in a different universality class than that of the original ground state $| \psi_0 \ra$, the system must necessarily pass through a QCP. In this Section, we estimate the optimal ramp speed, for which the system's state at time $t=0$ is as close as possible to the ground state of $H_{\rm eff}$. More precisely, we take the optimal ramp rate $s_*$ to be such that the combined heating effects due to both the drive and the fall-out of adiabaticity during the ramp are minimized. To that end, we find the scaling of the excess heat per unit volume $q_{\rm ramp}$ absorbed due to the non-adiabatic passage through the QCP as a result of the Kibble-Zurek mechanism analyzed through adiabatic perturbation theory \cite{PhysRevB.72.161201, PhysRevB.81.012303, 2010LNP...802...75D}.  We note that the results of this Section are sensitive to the choice of ramp profile, which is to be expected since the behavior of the system depends on how the critical point is traversed. We focus on the ramp profile $\lambda(t) = e^{st}$ that we have been considering in this paper; however, a similar analysis can be performed for other ramp profiles.

The general situation of interest is illustrated in Fig.~\ref{fig:function}: initially, the system $H_0$ is in a gapped phase 1. As the driving term is ramped, the effective Hamiltonian $H_{\rm eff}^{(s)}(t)$ crosses the QCP at a critical ramp value $\lambda_c$ at time $t_c$. This brings it into a  different gapped phase 2. These are illustrated by points (A), (B) and (C) respectively in Fig.~\ref{fig:function}. 
In what follows, we will focus on a slightly simpler scenario, where the static system, described by the Hamiltonian $H_0$, is already gapless (i.e. point (A) coincides with point (B)). Such a scenario is not pathological and was described e.g., in Ref.~\cite{PhysRevB.79.081406}, which proposed to realize the Haldane model in graphene (which is gapless at half-filling) by periodic driving which induces a Haldane mass gap.

Let us now understand how the effective Hamiltonian, initially at $H_0$, leaves the QCP as the drive is is turned on. From Eq.~(\ref{eqn:Expansion}, \ref{eqn:H0}), we see that the higher order terms $H^{(s,k)}(t)$, for $k \geq 1$, are the ones that `pull' the system away from criticality -- in renormalization group parlance, they are relevant operators to the fixed point. Furthermore, since the expansion is perturbative in $J/\omega$, we can concentrate on the effect of $H^{(s,1)}(t)$ to $H_0$, which we will assume is the dominant term driving the system away from  criticality. From Eq.~(\ref{eqn:firstOrder}), we see that there are two terms in $H^{(s,1)}(t)$,
\begin{align}
& \mathcal{O}_1 \equiv -i \sum_{n \neq 0}  \frac{[V_n,H]}{i \omega_n + s}  \text{ and } \nonumber \\
& \mathcal{O}_2 \equiv +\frac{i}{2} \sum_{n\neq 0} \frac{[V_n, V_{-n}]}{i\omega_n+s},
\end{align}
with scaling dimensions $\Delta_{1}$ and $\Delta_{2}$, that are being turned on at different rates $\lambda(t)$ and $\lambda(t)^2$ respectively. Note that $\mathcal{O}_2$ is being ramped up at a slower rate than $\mathcal{O}_1$.  We have therefore reduced the problem to understanding the slow quench dynamics of the following Hamiltonian
\be
H_0 + \lambda(t) \mathcal{O}_1 + \lambda(t)^2 \mathcal{O}_2,
\label{eqn:QCP}
\ee
where $\lambda(t) = e^{st}$. 

In general, the behavior of the system near the QCP, such as  the gap,  the correlation length, etc., will be a scaling function of critical exponents associated with both relevant operators. However, because of the difference in ramp speeds of $\mathcal{O}_1$ and $\mathcal{O}_2$, such quantities will generically instead be dominated by just one of the two relevant operators. Determining which operator dominates at the QCP involves comparing the scaling dimensions $(\Delta_1, \Delta_2)$ of the two operators and the ramp speeds $(\lambda, \lambda^2)$, and we leave a discussion of this determination to Appendix C. For the purposes of the remainder of this section, we only need the fact that there is generally only one dominant relevator operator; let us assume it is $\mathcal{O}_1$ with scaling dimension $\Delta = \Delta_1$, which is being ramped up as $\lambda(t) = e^{st}$. The result with $\mathcal{O}_2$ being the dominant relevant operator is identical with the replacement of all critical exponents to be those of $\mathcal{O}_2$'s and also $\lambda \to \lambda^2$.

Now, near the QCP, the system's gap $g$ vanishes as $g \sim \lambda^{z \nu}$, where $z$ is the dynamical critical exponent and $\nu$ the critical exponent of the correlation length $\xi \sim \lambda^{-\nu}$ associated with the dominant relevant operator. As a result, regardless of how slow  the ramp speed $s$ is, the system will not be adiabatic near the QCP, and it will fall out of equilibrium (i.e.~not remain in the instantaneous ground state). Nevertheless, there will be universality in the transition rates out of equilibrium because of the proximity to the QCP: to estimate the scaling of the excess heat per volume $q_{\rm ramp}$ created as an example, we have to identify a time-scale where the adiabaticity is violated ($dg/dt \sim g^2$) and then invoke general scaling arguments on certain response functions near the QCP. Since this is a well-studied topic \cite{PhysRevB.72.161201, PhysRevB.81.012303, 2010LNP...802...75D}, we simply adapt known results on the subject.

From adiabatic perturbation theory, with the system originally in the ground state at the beginning of the ramp, the transition amplitude from the instantaneous ground state $| 0 \rangle$ to an instantaneous excited state $| n\ra$ at the end of the ramp  is 
\begin{align}
\alpha_n(\lambda_f) \approx - \int_{\lambda_i }^{\lambda_f }  d\lambda' \la n(\lambda) | \partial_{\lambda'} | 0(\lambda)\ra e^{i(\Theta_n(t') - \Theta_0(t'))},
\label{eqn:alpha}
\end{align}
where $|n(\lambda)\ra$ are the instantaneous eigenstates of $H^{(s)}_{\rm eff}(\lambda)$, and the dynamical phase factor
\be 
\Theta_n(\lambda) = \int_{0}^{\lambda} d\lambda' \frac{E_n(\lambda')}{\dot{\lambda}'}.
\label{eqn:DPF}
\ee
The transition probability is therefore $|\alpha_n(\lambda_f)|^2$. 

To make more headway, let us further assume that we are dealing with a spatially homogenous system with quasiparticle excitations characterized by momenta $k$, and also that at the end of the ramp, the system is gapped with energy scale $\Delta_g(\omega)$ which is an algebraic function of the driving frequency, such as $\Delta_g(\omega) \sim J^2/\omega$. Then, $q_{\rm ramp}$ is roughly the gap multiplied by the density of quasiparticles created,
\begin{align}
q_{\rm ramp} \sim \Delta_g(\omega) n_{\rm ex} \approx \Delta_g(\omega) \int \frac{d^dk}{(2\pi)^d} |\alpha_k(\lambda_f)|^2.
\end{align}
To estimate this quantity, we note that for an exponential ramp the dynamical phase factor, Eqn.~(\ref{eqn:DPF}), can be written as
\begin{align}
\Theta_n(\lambda) = \frac{1}{s} \int_{0}^{\lambda}  d \lambda' \lambda'^{-1} (\epsilon_k(\lambda') - \epsilon_0(\lambda')).
\label{eqn:DPF2}
\end{align}
Following Refs.~\cite{PhysRevB.72.161201, PhysRevB.81.012303, 2010LNP...802...75D}, near the QCP the quasiparticle energy can be written using a scaling function $F$:
\begin{align}
\epsilon_k(\lambda') - \epsilon_0(\lambda') = \lambda^{z \nu} F(k/\lambda^\nu),
\label{eqn:Fscaling}
\end{align}
which has asymptotics $F(x) \propto x^z$ for $x \gg 1$, and $F(x) \to \text{const}$ as $x \to 0$. The scaling, Eqn.~(\ref{eqn:Fscaling}), put into Eqn.~(\ref{eqn:DPF2}), suggests the change of variables
\begin{align}
\lambda = \xi s^{\frac{1}{\nu z}},  ~~~ k = \eta s^{\frac{1}{z}}.
\end{align}
Adopting another scaling ansatz for the matrix element,
\begin{align}
\la k | \partial_\lambda |0 \ra = -\frac{\la k | V | 0 \ra}{\epsilon_k(\lambda) - \epsilon_0(\lambda) } = \frac{1}{\lambda} G(k/\lambda^\nu),
\end{align}
with asymptotics $G(x) \propto x^{-1/\nu}$ for $x \gg 1$, while $G(x) \propto x^\beta$ for $x \ll 1$, where $\beta$ is some non-negative number.

Then,
\begin{align}
n_{\rm ex} \approx s^{\frac{d}{z}} \int \frac{d^d \eta}{(2\pi)^d} |\alpha(\eta)|^2,
\end{align}
where the integral gives a number of order 1 if the dimensionality $d$ is below some upper critical dimesion \cite{PhysRevB.72.161201, PhysRevB.81.012303, 2010LNP...802...75D}.  

Thus, we have the result that the excess heat per unit volume due to an exponential ramp characterized by a ramp speed $s$ is
\begin{align}
q_{\rm ramp} \sim \Delta_g(\omega) s^{\frac{d}{z}}.
\end{align}

{\it Optimal ramp rate $s_*$.} We are now in a position to estimate the optimal rap rate $s_*$ by minizming the heating effects due to both the drive and the ramp. The local heating due to the drive as measured by $H_{\rm eff}^{(s)}(0)$ can be estimated from Eqn.~(\ref{eqn:Odiff}), by setting $O$ to be the energy density operator. We have
\begin{align}
q_{\rm drive} \sim e^{-\mathcal{C} \frac{\omega}{J}} s^{-2}.
\end{align}
Then, the total heating per unit volume is 
\begin{align}
q_{\rm drive} + q_{\rm ramp} \sim \Delta_g(\omega) s^{\frac{d}{z}} + e^{-\mathcal{C} \frac{\omega}{J}} s^{-2}.
\end{align}
Minimizing with respect to $s$, we find the optimal ramp rate
\begin{align}
s_* \approx \left( \frac{2 z}{d \Delta_g(\omega)} \right)^{\frac{z}{d+2z}} e^{- \frac{z}{d+2z} \mathcal{C} \frac{\omega}{J}} \sim t_*^{-\frac{z}{d+2z}}.
\end{align}
Since the asymptotic behavior as  $\omega \to \infty$ is governed by the exponential, we see that the optimal ramp rate should be chosen to be exponentially small in driving frequency but as some power of the inverse of the characteristic heating time-scale. At this optimal ramp rate $s_*$, both heating effects $q_{\rm drive} + q_{\rm ramp}$ are then exponentially small.

\section{Summary and discussion}
\label{sec:conclusion}

To summarize, in this paper we have developed a theoretical framework for describing the dynamics of Floquet many-body systems under slow ramping up of the drive. Our central result is that, provided the ramp time scale is short compared to the heating time scale, $s \gg t_*^{-1} \sim e^{-\mathcal{C} \frac{\omega}{J}}$ (condition (\ref{eqn:condition_s})), the dynamics of the system is well-approximated by the unitary evolution with an effective, slowly varying Hamiltonian $H_{\rm eff}^{(s)}(t)$. To arrive at this result, we performed a series of quasi-local, unitary transformations which were chosen to decouple slow and fast motion of the system. This allowed us to represent the transformed Hamiltonian as the sum of two parts: $H_{\rm eff}^{(s)}(t)$, which varies on the time scale $s^{-1}$, responsible for the slow-modulation dynamics, and an exponentially small, rapidly oscillating term whose norm is on the order $e^{-\mathcal{C} \frac{\omega}{J}}$, which describes the slow energy absorption from the drive. The condition (\ref{eqn:condition_s}) means that the ramp is done sufficiently fast such that the rapidly oscillating term in the Hamiltonian is negligible and does not lead to sizeable heating of the system.

The above result is general, and holds for any initial state of the system, and also for ramping the driving amplitude between two arbitrary values. Perhaps the most interesting application of this result it that it can be used to develop optimal strategies for preparing the desired many-body states in Floquet systems. Previous works had proposed an ``adiabatic'' preparation of the ground state of $H_{\rm eff}$ (the target state): starting from the ground state of a static Hamiltonian which can be easily prepared in experiment, subsequently slowly ramping up the drive to a value which gives the desired effective Hamiltonian $H_{\rm eff}$. Our work shows that, generally, no adiabatic limit exists, since, if $s < e^{-\mathcal{C} \frac{\omega}{J}}$, the system will significantly heat up, ending up far from the ground state of $H_{\rm eff}$. However, if condition (\ref{eqn:condition_s}) is met, the system  can end up close to the ground state of $H_{\rm eff}$, because now $H_{\rm eff}^{(s)}(t)$ describes a  slow quench from $H_0$ to $H_{\rm eff}$.

Furthermore, in general, we argued that if we hope to prepare a Floquet `ground state' in a different universality class than that of the initial state, the ramp must inevitably go through a QCP. For this scenario, we found an optimal ramp rate which minimizes both the heating effects due to the drive and the lack of adiabaticity as the system crosses the QCP, by using adiabatic perturbation theory and general scaling arguments to extract the universal behavior of the excess heat per unit volume. We found that the optimal ramp rate should also be chosen exponentially small in frequency, but as some power of the inverse of characteristic heating time-scale, at which point both heating effects are exponentially minimized. Thus, it is possible to prepare the `ground state' of Floquet systems with a high degree of accuracy and control. 

Our results complement the proposals  to use Floquet engineering to realize interesting and exotic phases of matter, such as symmetry protected topological phases (SPTs) or fractional Chern insulators (FCIs). While there is now a panoply of such proposals, there has been a relative lack of theoretical work regarding the preparation of the desired states in these systems. Our work here fills this void and bridges the gap between theory and experiment, potentially paving the way for the experimental realization of a rich variety of Floquet many-body states. 

Finally, it is interesting to note that a similar problem was previously studied for {\it single-particle} systems with an infinite number of levels by Hone, Ketzmerick and Kohn~\cite{Kohn97}, who showed that such systems do not have a well-defined adiabatic limit. While this conclusion is similar to what we found for many-body systems, the physics is quite different. In the single-particle case, adiabatic dynamics breaks down because of the localization of Floquet eigenstates in the energy space, which leads to level crossings with arbitrarily small gaps. In contrast, in many-body systems, the adiabatic limit breaks down because of inevitable heating effects, which arise because many-body Floquet eigenstates are ergodic rather than localized.

\section{Acknowledgments}
W.~W.~H.~thanks Mark Spencer Rudner and Max Metlitski for helpful comments and enlightening discussions. D.~A.~A.~ and W.~W.~H.~thank the Kavli Institute for Theoretical Physics in Santa Barbara, at which a large part of the work was done, for their hospitality. This research was supported by Swiss National Science Foundation, and in part by the National Science Foundation under Grant No.~NSF PHY11-25915. 

\bibliography{refs}

\begin{thebibliography}{43}%
\makeatletter
\providecommand \@ifxundefined [1]{%
 \@ifx{#1\undefined}
}%
\providecommand \@ifnum [1]{%
 \ifnum #1\expandafter \@firstoftwo
 \else \expandafter \@secondoftwo
 \fi
}%
\providecommand \@ifx [1]{%
 \ifx #1\expandafter \@firstoftwo
 \else \expandafter \@secondoftwo
 \fi
}%
\providecommand \natexlab [1]{#1}%
\providecommand \enquote  [1]{``#1''}%
\providecommand \bibnamefont  [1]{#1}%
\providecommand \bibfnamefont [1]{#1}%
\providecommand \citenamefont [1]{#1}%
\providecommand \href@noop [0]{\@secondoftwo}%
\providecommand \href [0]{\begingroup \@sanitize@url \@href}%
\providecommand \@href[1]{\@@startlink{#1}\@@href}%
\providecommand \@@href[1]{\endgroup#1\@@endlink}%
\providecommand \@sanitize@url [0]{\catcode `\\12\catcode `\$12\catcode
  `\&12\catcode `\#12\catcode `\^12\catcode `\_12\catcode `\%12\relax}%
\providecommand \@@startlink[1]{}%
\providecommand \@@endlink[0]{}%
\providecommand \url  [0]{\begingroup\@sanitize@url \@url }%
\providecommand \@url [1]{\endgroup\@href {#1}{\urlprefix }}%
\providecommand \urlprefix  [0]{URL }%
\providecommand \Eprint [0]{\href }%
\providecommand \doibase [0]{http://dx.doi.org/}%
\providecommand \selectlanguage [0]{\@gobble}%
\providecommand \bibinfo  [0]{\@secondoftwo}%
\providecommand \bibfield  [0]{\@secondoftwo}%
\providecommand \translation [1]{[#1]}%
\providecommand \BibitemOpen [0]{}%
\providecommand \bibitemStop [0]{}%
\providecommand \bibitemNoStop [0]{.\EOS\space}%
\providecommand \EOS [0]{\spacefactor3000\relax}%
\providecommand \BibitemShut  [1]{\csname bibitem#1\endcsname}%
\let\auto@bib@innerbib\@empty
\bibitem [{\citenamefont {Lignier}\ \emph {et~al.}(2007)\citenamefont
  {Lignier}, \citenamefont {Sias}, \citenamefont {Ciampini}, \citenamefont
  {Singh}, \citenamefont {Zenesini}, \citenamefont {Morsch},\ and\
  \citenamefont {Arimondo}}]{Arimondo07}%
  \BibitemOpen
  \bibfield  {author} {\bibinfo {author} {\bibfnamefont {H.}~\bibnamefont
  {Lignier}}, \bibinfo {author} {\bibfnamefont {C.}~\bibnamefont {Sias}},
  \bibinfo {author} {\bibfnamefont {D.}~\bibnamefont {Ciampini}}, \bibinfo
  {author} {\bibfnamefont {Y.}~\bibnamefont {Singh}}, \bibinfo {author}
  {\bibfnamefont {A.}~\bibnamefont {Zenesini}}, \bibinfo {author}
  {\bibfnamefont {O.}~\bibnamefont {Morsch}}, \ and\ \bibinfo {author}
  {\bibfnamefont {E.}~\bibnamefont {Arimondo}},\ }\href {\doibase
  10.1103/PhysRevLett.99.220403} {\bibfield  {journal} {\bibinfo  {journal}
  {Phys. Rev. Lett.}\ }\textbf {\bibinfo {volume} {99}},\ \bibinfo {pages}
  {220403} (\bibinfo {year} {2007})}\BibitemShut {NoStop}%
\bibitem [{\citenamefont {Kierig}\ \emph {et~al.}(2008)\citenamefont {Kierig},
  \citenamefont {Schnorrberger}, \citenamefont {Schietinger}, \citenamefont
  {Tomkovic},\ and\ \citenamefont {Oberthaler}}]{Obertaler08}%
  \BibitemOpen
  \bibfield  {author} {\bibinfo {author} {\bibfnamefont {E.}~\bibnamefont
  {Kierig}}, \bibinfo {author} {\bibfnamefont {U.}~\bibnamefont
  {Schnorrberger}}, \bibinfo {author} {\bibfnamefont {A.}~\bibnamefont
  {Schietinger}}, \bibinfo {author} {\bibfnamefont {J.}~\bibnamefont
  {Tomkovic}}, \ and\ \bibinfo {author} {\bibfnamefont {M.~K.}\ \bibnamefont
  {Oberthaler}},\ }\href {\doibase 10.1103/PhysRevLett.100.190405} {\bibfield
  {journal} {\bibinfo  {journal} {Phys. Rev. Lett.}\ }\textbf {\bibinfo
  {volume} {100}},\ \bibinfo {pages} {190405} (\bibinfo {year}
  {2008})}\BibitemShut {NoStop}%
\bibitem [{\citenamefont {Zenesini}\ \emph {et~al.}(2009)\citenamefont
  {Zenesini}, \citenamefont {Lignier}, \citenamefont {Ciampini}, \citenamefont
  {Morsch},\ and\ \citenamefont {Arimondo}}]{Arimondo09}%
  \BibitemOpen
  \bibfield  {author} {\bibinfo {author} {\bibfnamefont {A.}~\bibnamefont
  {Zenesini}}, \bibinfo {author} {\bibfnamefont {H.}~\bibnamefont {Lignier}},
  \bibinfo {author} {\bibfnamefont {D.}~\bibnamefont {Ciampini}}, \bibinfo
  {author} {\bibfnamefont {O.}~\bibnamefont {Morsch}}, \ and\ \bibinfo {author}
  {\bibfnamefont {E.}~\bibnamefont {Arimondo}},\ }\href {\doibase
  10.1103/PhysRevLett.102.100403} {\bibfield  {journal} {\bibinfo  {journal}
  {Phys. Rev. Lett.}\ }\textbf {\bibinfo {volume} {102}},\ \bibinfo {pages}
  {100403} (\bibinfo {year} {2009})}\BibitemShut {NoStop}%
\bibitem [{\citenamefont {Aidelsburger}\ \emph {et~al.}(2013)\citenamefont
  {Aidelsburger}, \citenamefont {Atala}, \citenamefont {Lohse}, \citenamefont
  {Barreiro}, \citenamefont {Paredes},\ and\ \citenamefont
  {Bloch}}]{PhysRevLett.111.185301}%
  \BibitemOpen
  \bibfield  {author} {\bibinfo {author} {\bibfnamefont {M.}~\bibnamefont
  {Aidelsburger}}, \bibinfo {author} {\bibfnamefont {M.}~\bibnamefont {Atala}},
  \bibinfo {author} {\bibfnamefont {M.}~\bibnamefont {Lohse}}, \bibinfo
  {author} {\bibfnamefont {J.~T.}\ \bibnamefont {Barreiro}}, \bibinfo {author}
  {\bibfnamefont {B.}~\bibnamefont {Paredes}}, \ and\ \bibinfo {author}
  {\bibfnamefont {I.}~\bibnamefont {Bloch}},\ }\href {\doibase
  10.1103/PhysRevLett.111.185301} {\bibfield  {journal} {\bibinfo  {journal}
  {Phys. Rev. Lett.}\ }\textbf {\bibinfo {volume} {111}},\ \bibinfo {pages}
  {185301} (\bibinfo {year} {2013})}\BibitemShut {NoStop}%
\bibitem [{\citenamefont {Eckardt}\ \emph {et~al.}(2005)\citenamefont
  {Eckardt}, \citenamefont {Weiss},\ and\ \citenamefont
  {Holthaus}}]{Eckardt05}%
  \BibitemOpen
  \bibfield  {author} {\bibinfo {author} {\bibfnamefont {A.}~\bibnamefont
  {Eckardt}}, \bibinfo {author} {\bibfnamefont {C.}~\bibnamefont {Weiss}}, \
  and\ \bibinfo {author} {\bibfnamefont {M.}~\bibnamefont {Holthaus}},\ }\href
  {\doibase 10.1103/PhysRevLett.95.260404} {\bibfield  {journal} {\bibinfo
  {journal} {Phys. Rev. Lett.}\ }\textbf {\bibinfo {volume} {95}},\ \bibinfo
  {pages} {260404} (\bibinfo {year} {2005})}\BibitemShut {NoStop}%
\bibitem [{\citenamefont {Oka}\ and\ \citenamefont
  {Aoki}(2009)}]{PhysRevB.79.081406}%
  \BibitemOpen
  \bibfield  {author} {\bibinfo {author} {\bibfnamefont {T.}~\bibnamefont
  {Oka}}\ and\ \bibinfo {author} {\bibfnamefont {H.}~\bibnamefont {Aoki}},\
  }\href {\doibase 10.1103/PhysRevB.79.081406} {\bibfield  {journal} {\bibinfo
  {journal} {Phys. Rev. B}\ }\textbf {\bibinfo {volume} {79}},\ \bibinfo
  {pages} {081406} (\bibinfo {year} {2009})}\BibitemShut {NoStop}%
\bibitem [{\citenamefont {Kitagawa}\ \emph {et~al.}(2010)\citenamefont
  {Kitagawa}, \citenamefont {Berg}, \citenamefont {Rudner},\ and\ \citenamefont
  {Demler}}]{PhysRevB.82.235114}%
  \BibitemOpen
  \bibfield  {author} {\bibinfo {author} {\bibfnamefont {T.}~\bibnamefont
  {Kitagawa}}, \bibinfo {author} {\bibfnamefont {E.}~\bibnamefont {Berg}},
  \bibinfo {author} {\bibfnamefont {M.}~\bibnamefont {Rudner}}, \ and\ \bibinfo
  {author} {\bibfnamefont {E.}~\bibnamefont {Demler}},\ }\href {\doibase
  10.1103/PhysRevB.82.235114} {\bibfield  {journal} {\bibinfo  {journal} {Phys.
  Rev. B}\ }\textbf {\bibinfo {volume} {82}},\ \bibinfo {pages} {235114}
  (\bibinfo {year} {2010})}\BibitemShut {NoStop}%
\bibitem [{\citenamefont {Eckardt}\ \emph {et~al.}(2010)\citenamefont
  {Eckardt}, \citenamefont {Hauke}, \citenamefont {Soltan-Panahi},
  \citenamefont {Becker}, \citenamefont {Sengstock},\ and\ \citenamefont
  {Lewenstein}}]{Eckardt10}%
  \BibitemOpen
  \bibfield  {author} {\bibinfo {author} {\bibfnamefont {A.}~\bibnamefont
  {Eckardt}}, \bibinfo {author} {\bibfnamefont {P.}~\bibnamefont {Hauke}},
  \bibinfo {author} {\bibfnamefont {P.}~\bibnamefont {Soltan-Panahi}}, \bibinfo
  {author} {\bibfnamefont {C.}~\bibnamefont {Becker}}, \bibinfo {author}
  {\bibfnamefont {K.}~\bibnamefont {Sengstock}}, \ and\ \bibinfo {author}
  {\bibfnamefont {M.}~\bibnamefont {Lewenstein}},\ }\href
  {http://stacks.iop.org/0295-5075/89/i=1/a=10010} {\bibfield  {journal}
  {\bibinfo  {journal} {EPL (Europhysics Letters)}\ }\textbf {\bibinfo {volume}
  {89}},\ \bibinfo {pages} {10010} (\bibinfo {year} {2010})}\BibitemShut
  {NoStop}%
\bibitem [{\citenamefont {{Lindner}}\ \emph {et~al.}(2011)\citenamefont
  {{Lindner}}, \citenamefont {{Refael}},\ and\ \citenamefont
  {{Galitski}}}]{2011NatPh...7..490L}%
  \BibitemOpen
  \bibfield  {author} {\bibinfo {author} {\bibfnamefont {N.~H.}\ \bibnamefont
  {{Lindner}}}, \bibinfo {author} {\bibfnamefont {G.}~\bibnamefont {{Refael}}},
  \ and\ \bibinfo {author} {\bibfnamefont {V.}~\bibnamefont {{Galitski}}},\
  }\href {\doibase 10.1038/nphys1926} {\bibfield  {journal} {\bibinfo
  {journal} {Nature Physics}\ }\textbf {\bibinfo {volume} {7}},\ \bibinfo
  {pages} {490} (\bibinfo {year} {2011})},\ \Eprint
  {http://arxiv.org/abs/1008.1792} {arXiv:1008.1792 [cond-mat.mtrl-sci]}
  \BibitemShut {NoStop}%
\bibitem [{\citenamefont {Poletti}\ and\ \citenamefont
  {Kollath}(2011)}]{Kollath11}%
  \BibitemOpen
  \bibfield  {author} {\bibinfo {author} {\bibfnamefont {D.}~\bibnamefont
  {Poletti}}\ and\ \bibinfo {author} {\bibfnamefont {C.}~\bibnamefont
  {Kollath}},\ }\href {\doibase 10.1103/PhysRevA.84.013615} {\bibfield
  {journal} {\bibinfo  {journal} {Phys. Rev. A}\ }\textbf {\bibinfo {volume}
  {84}},\ \bibinfo {pages} {013615} (\bibinfo {year} {2011})}\BibitemShut
  {NoStop}%
\bibitem [{\citenamefont {Kitagawa}\ \emph {et~al.}(2011)\citenamefont
  {Kitagawa}, \citenamefont {Oka}, \citenamefont {Brataas}, \citenamefont
  {Fu},\ and\ \citenamefont {Demler}}]{Kitagawa11}%
  \BibitemOpen
  \bibfield  {author} {\bibinfo {author} {\bibfnamefont {T.}~\bibnamefont
  {Kitagawa}}, \bibinfo {author} {\bibfnamefont {T.}~\bibnamefont {Oka}},
  \bibinfo {author} {\bibfnamefont {A.}~\bibnamefont {Brataas}}, \bibinfo
  {author} {\bibfnamefont {L.}~\bibnamefont {Fu}}, \ and\ \bibinfo {author}
  {\bibfnamefont {E.}~\bibnamefont {Demler}},\ }\href {\doibase
  10.1103/PhysRevB.84.235108} {\bibfield  {journal} {\bibinfo  {journal} {Phys.
  Rev. B}\ }\textbf {\bibinfo {volume} {84}},\ \bibinfo {pages} {235108}
  (\bibinfo {year} {2011})}\BibitemShut {NoStop}%
\bibitem [{\citenamefont {Rudner}\ \emph {et~al.}(2013)\citenamefont {Rudner},
  \citenamefont {Lindner}, \citenamefont {Berg},\ and\ \citenamefont
  {Levin}}]{PhysRevX.3.031005}%
  \BibitemOpen
  \bibfield  {author} {\bibinfo {author} {\bibfnamefont {M.~S.}\ \bibnamefont
  {Rudner}}, \bibinfo {author} {\bibfnamefont {N.~H.}\ \bibnamefont {Lindner}},
  \bibinfo {author} {\bibfnamefont {E.}~\bibnamefont {Berg}}, \ and\ \bibinfo
  {author} {\bibfnamefont {M.}~\bibnamefont {Levin}},\ }\href {\doibase
  10.1103/PhysRevX.3.031005} {\bibfield  {journal} {\bibinfo  {journal} {Phys.
  Rev. X}\ }\textbf {\bibinfo {volume} {3}},\ \bibinfo {pages} {031005}
  (\bibinfo {year} {2013})}\BibitemShut {NoStop}%
\bibitem [{\citenamefont {Iadecola}\ \emph {et~al.}(2015)\citenamefont
  {Iadecola}, \citenamefont {Santos},\ and\ \citenamefont
  {Chamon}}]{PhysRevB.92.125107}%
  \BibitemOpen
  \bibfield  {author} {\bibinfo {author} {\bibfnamefont {T.}~\bibnamefont
  {Iadecola}}, \bibinfo {author} {\bibfnamefont {L.~H.}\ \bibnamefont
  {Santos}}, \ and\ \bibinfo {author} {\bibfnamefont {C.}~\bibnamefont
  {Chamon}},\ }\href {\doibase 10.1103/PhysRevB.92.125107} {\bibfield
  {journal} {\bibinfo  {journal} {Phys. Rev. B}\ }\textbf {\bibinfo {volume}
  {92}},\ \bibinfo {pages} {125107} (\bibinfo {year} {2015})}\BibitemShut
  {NoStop}%
\bibitem [{\citenamefont {Goldman}\ and\ \citenamefont
  {Dalibard}(2014)}]{Goldman14}%
  \BibitemOpen
  \bibfield  {author} {\bibinfo {author} {\bibfnamefont {N.}~\bibnamefont
  {Goldman}}\ and\ \bibinfo {author} {\bibfnamefont {J.}~\bibnamefont
  {Dalibard}},\ }\href {\doibase 10.1103/PhysRevX.4.031027} {\bibfield
  {journal} {\bibinfo  {journal} {Phys. Rev. X}\ }\textbf {\bibinfo {volume}
  {4}},\ \bibinfo {pages} {031027} (\bibinfo {year} {2014})}\BibitemShut
  {NoStop}%
\bibitem [{\citenamefont {Eckardt}\ and\ \citenamefont
  {Anisimovas}(2015)}]{1367-2630-17-9-093039}%
  \BibitemOpen
  \bibfield  {author} {\bibinfo {author} {\bibfnamefont {A.}~\bibnamefont
  {Eckardt}}\ and\ \bibinfo {author} {\bibfnamefont {E.}~\bibnamefont
  {Anisimovas}},\ }\href {http://stacks.iop.org/1367-2630/17/i=9/a=093039}
  {\bibfield  {journal} {\bibinfo  {journal} {New Journal of Physics}\ }\textbf
  {\bibinfo {volume} {17}},\ \bibinfo {pages} {093039} (\bibinfo {year}
  {2015})}\BibitemShut {NoStop}%
\bibitem [{\citenamefont {Bloch}\ \emph {et~al.}(2008)\citenamefont {Bloch},
  \citenamefont {Dalibard},\ and\ \citenamefont {Zwerger}}]{RevModPhys.80.885}%
  \BibitemOpen
  \bibfield  {author} {\bibinfo {author} {\bibfnamefont {I.}~\bibnamefont
  {Bloch}}, \bibinfo {author} {\bibfnamefont {J.}~\bibnamefont {Dalibard}}, \
  and\ \bibinfo {author} {\bibfnamefont {W.}~\bibnamefont {Zwerger}},\ }\href
  {\doibase 10.1103/RevModPhys.80.885} {\bibfield  {journal} {\bibinfo
  {journal} {Rev. Mod. Phys.}\ }\textbf {\bibinfo {volume} {80}},\ \bibinfo
  {pages} {885} (\bibinfo {year} {2008})}\BibitemShut {NoStop}%
\bibitem [{\citenamefont {{Jotzu}}\ \emph {et~al.}(2014)\citenamefont
  {{Jotzu}}, \citenamefont {{Messer}}, \citenamefont {{Desbuquois}},
  \citenamefont {{Lebrat}}, \citenamefont {{Uehlinger}}, \citenamefont
  {{Greif}},\ and\ \citenamefont {{Esslinger}}}]{2014Natur.515..237J}%
  \BibitemOpen
  \bibfield  {author} {\bibinfo {author} {\bibfnamefont {G.}~\bibnamefont
  {{Jotzu}}}, \bibinfo {author} {\bibfnamefont {M.}~\bibnamefont {{Messer}}},
  \bibinfo {author} {\bibfnamefont {R.}~\bibnamefont {{Desbuquois}}}, \bibinfo
  {author} {\bibfnamefont {M.}~\bibnamefont {{Lebrat}}}, \bibinfo {author}
  {\bibfnamefont {T.}~\bibnamefont {{Uehlinger}}}, \bibinfo {author}
  {\bibfnamefont {D.}~\bibnamefont {{Greif}}}, \ and\ \bibinfo {author}
  {\bibfnamefont {T.}~\bibnamefont {{Esslinger}}},\ }\href {\doibase
  10.1038/nature13915} {\bibfield  {journal} {\bibinfo  {journal} {\nat}\
  }\textbf {\bibinfo {volume} {515}},\ \bibinfo {pages} {237} (\bibinfo {year}
  {2014})},\ \Eprint {http://arxiv.org/abs/1406.7874} {arXiv:1406.7874
  [cond-mat.quant-gas]} \BibitemShut {NoStop}%
\bibitem [{\citenamefont {{Aidelsburger}}\ \emph {et~al.}(2015)\citenamefont
  {{Aidelsburger}}, \citenamefont {{Lohse}}, \citenamefont {{Schweizer}},
  \citenamefont {{Atala}}, \citenamefont {{Barreiro}}, \citenamefont
  {{Nascimb{\`e}ne}}, \citenamefont {{Cooper}}, \citenamefont {{Bloch}},\ and\
  \citenamefont {{Goldman}}}]{2015NatPh..11..162A}%
  \BibitemOpen
  \bibfield  {author} {\bibinfo {author} {\bibfnamefont {M.}~\bibnamefont
  {{Aidelsburger}}}, \bibinfo {author} {\bibfnamefont {M.}~\bibnamefont
  {{Lohse}}}, \bibinfo {author} {\bibfnamefont {C.}~\bibnamefont
  {{Schweizer}}}, \bibinfo {author} {\bibfnamefont {M.}~\bibnamefont
  {{Atala}}}, \bibinfo {author} {\bibfnamefont {J.~T.}\ \bibnamefont
  {{Barreiro}}}, \bibinfo {author} {\bibfnamefont {S.}~\bibnamefont
  {{Nascimb{\`e}ne}}}, \bibinfo {author} {\bibfnamefont {N.~R.}\ \bibnamefont
  {{Cooper}}}, \bibinfo {author} {\bibfnamefont {I.}~\bibnamefont {{Bloch}}}, \
  and\ \bibinfo {author} {\bibfnamefont {N.}~\bibnamefont {{Goldman}}},\ }\href
  {\doibase 10.1038/nphys3171} {\bibfield  {journal} {\bibinfo  {journal}
  {Nature Physics}\ }\textbf {\bibinfo {volume} {11}},\ \bibinfo {pages} {162}
  (\bibinfo {year} {2015})},\ \Eprint {http://arxiv.org/abs/1407.4205}
  {arXiv:1407.4205 [cond-mat.quant-gas]} \BibitemShut {NoStop}%
\bibitem [{\citenamefont {Grushin}\ \emph {et~al.}(2014)\citenamefont
  {Grushin}, \citenamefont {G\'omez-Le\'on},\ and\ \citenamefont
  {Neupert}}]{PhysRevLett.112.156801}%
  \BibitemOpen
  \bibfield  {author} {\bibinfo {author} {\bibfnamefont {A.~G.}\ \bibnamefont
  {Grushin}}, \bibinfo {author} {\bibfnamefont {A.}~\bibnamefont
  {G\'omez-Le\'on}}, \ and\ \bibinfo {author} {\bibfnamefont {T.}~\bibnamefont
  {Neupert}},\ }\href {\doibase 10.1103/PhysRevLett.112.156801} {\bibfield
  {journal} {\bibinfo  {journal} {Phys. Rev. Lett.}\ }\textbf {\bibinfo
  {volume} {112}},\ \bibinfo {pages} {156801} (\bibinfo {year}
  {2014})}\BibitemShut {NoStop}%
\bibitem [{\citenamefont {{Abanin}}\ \emph
  {et~al.}(2015{\natexlab{a}})\citenamefont {{Abanin}}, \citenamefont {{De
  Roeck}}, \citenamefont {{Huveneers}},\ and\ \citenamefont
  {{Ho}}}]{2015arXiv150905386A}%
  \BibitemOpen
  \bibfield  {author} {\bibinfo {author} {\bibfnamefont {D.}~\bibnamefont
  {{Abanin}}}, \bibinfo {author} {\bibfnamefont {W.}~\bibnamefont {{De
  Roeck}}}, \bibinfo {author} {\bibfnamefont {F.}~\bibnamefont {{Huveneers}}},
  \ and\ \bibinfo {author} {\bibfnamefont {W.~W.}\ \bibnamefont {{Ho}}},\
  }\href@noop {} {\bibfield  {journal} {\bibinfo  {journal} {ArXiv e-prints}\ }
  (\bibinfo {year} {2015}{\natexlab{a}})},\ \Eprint
  {http://arxiv.org/abs/1509.05386} {arXiv:1509.05386 [math-ph]} \BibitemShut
  {NoStop}%
\bibitem [{\citenamefont {{Abanin}}\ \emph
  {et~al.}(2015{\natexlab{b}})\citenamefont {{Abanin}}, \citenamefont {{De
  Roeck}},\ and\ \citenamefont {{Ho}}}]{2015arXiv151003405A}%
  \BibitemOpen
  \bibfield  {author} {\bibinfo {author} {\bibfnamefont {D.~A.}\ \bibnamefont
  {{Abanin}}}, \bibinfo {author} {\bibfnamefont {W.}~\bibnamefont {{De
  Roeck}}}, \ and\ \bibinfo {author} {\bibfnamefont {W.~W.}\ \bibnamefont
  {{Ho}}},\ }\href@noop {} {\bibfield  {journal} {\bibinfo  {journal} {ArXiv
  e-prints}\ } (\bibinfo {year} {2015}{\natexlab{b}})},\ \Eprint
  {http://arxiv.org/abs/1510.03405} {arXiv:1510.03405 [cond-mat.stat-mech]}
  \BibitemShut {NoStop}%
\bibitem [{\citenamefont {Mori}\ \emph {et~al.}(2016)\citenamefont {Mori},
  \citenamefont {Kuwahara},\ and\ \citenamefont
  {Saito}}]{PhysRevLett.116.120401}%
  \BibitemOpen
  \bibfield  {author} {\bibinfo {author} {\bibfnamefont {T.}~\bibnamefont
  {Mori}}, \bibinfo {author} {\bibfnamefont {T.}~\bibnamefont {Kuwahara}}, \
  and\ \bibinfo {author} {\bibfnamefont {K.}~\bibnamefont {Saito}},\ }\href
  {\doibase 10.1103/PhysRevLett.116.120401} {\bibfield  {journal} {\bibinfo
  {journal} {Phys. Rev. Lett.}\ }\textbf {\bibinfo {volume} {116}},\ \bibinfo
  {pages} {120401} (\bibinfo {year} {2016})}\BibitemShut {NoStop}%
\bibitem [{\citenamefont {Kuwahara}\ \emph {et~al.}(2016)\citenamefont
  {Kuwahara}, \citenamefont {Mori},\ and\ \citenamefont
  {Saito}}]{Kuwahara201696}%
  \BibitemOpen
  \bibfield  {author} {\bibinfo {author} {\bibfnamefont {T.}~\bibnamefont
  {Kuwahara}}, \bibinfo {author} {\bibfnamefont {T.}~\bibnamefont {Mori}}, \
  and\ \bibinfo {author} {\bibfnamefont {K.}~\bibnamefont {Saito}},\ }\href
  {\doibase http://dx.doi.org/10.1016/j.aop.2016.01.012} {\bibfield  {journal}
  {\bibinfo  {journal} {Annals of Physics}\ }\textbf {\bibinfo {volume}
  {367}},\ \bibinfo {pages} {96 } (\bibinfo {year} {2016})}\BibitemShut
  {NoStop}%
\bibitem [{\citenamefont {Blanes}\ \emph {et~al.}(2009)\citenamefont {Blanes},
  \citenamefont {Casas}, \citenamefont {Oteo},\ and\ \citenamefont
  {Ros}}]{Blanes2009151}%
  \BibitemOpen
  \bibfield  {author} {\bibinfo {author} {\bibfnamefont {S.}~\bibnamefont
  {Blanes}}, \bibinfo {author} {\bibfnamefont {F.}~\bibnamefont {Casas}},
  \bibinfo {author} {\bibfnamefont {J.}~\bibnamefont {Oteo}}, \ and\ \bibinfo
  {author} {\bibfnamefont {J.}~\bibnamefont {Ros}},\ }\href {\doibase
  http://dx.doi.org/10.1016/j.physrep.2008.11.001} {\bibfield  {journal}
  {\bibinfo  {journal} {Physics Reports}\ }\textbf {\bibinfo {volume} {470}},\
  \bibinfo {pages} {151 } (\bibinfo {year} {2009})}\BibitemShut {NoStop}%
\bibitem [{\citenamefont {{Bukov}}\ \emph {et~al.}(2015)\citenamefont
  {{Bukov}}, \citenamefont {{D'Alessio}},\ and\ \citenamefont
  {{Polkovnikov}}}]{2015AdPhy..64..139B}%
  \BibitemOpen
  \bibfield  {author} {\bibinfo {author} {\bibfnamefont {M.}~\bibnamefont
  {{Bukov}}}, \bibinfo {author} {\bibfnamefont {L.}~\bibnamefont
  {{D'Alessio}}}, \ and\ \bibinfo {author} {\bibfnamefont {A.}~\bibnamefont
  {{Polkovnikov}}},\ }\href {\doibase 10.1080/00018732.2015.1055918} {\bibfield
   {journal} {\bibinfo  {journal} {Advances in Physics}\ }\textbf {\bibinfo
  {volume} {64}},\ \bibinfo {pages} {139} (\bibinfo {year} {2015})},\ \Eprint
  {http://arxiv.org/abs/1407.4803} {arXiv:1407.4803 [cond-mat.quant-gas]}
  \BibitemShut {NoStop}%
\bibitem [{\citenamefont {Lazarides}\ \emph {et~al.}(2014)\citenamefont
  {Lazarides}, \citenamefont {Das},\ and\ \citenamefont
  {Moessner}}]{PhysRevE.90.012110}%
  \BibitemOpen
  \bibfield  {author} {\bibinfo {author} {\bibfnamefont {A.}~\bibnamefont
  {Lazarides}}, \bibinfo {author} {\bibfnamefont {A.}~\bibnamefont {Das}}, \
  and\ \bibinfo {author} {\bibfnamefont {R.}~\bibnamefont {Moessner}},\ }\href
  {\doibase 10.1103/PhysRevE.90.012110} {\bibfield  {journal} {\bibinfo
  {journal} {Phys. Rev. E}\ }\textbf {\bibinfo {volume} {90}},\ \bibinfo
  {pages} {012110} (\bibinfo {year} {2014})}\BibitemShut {NoStop}%
\bibitem [{\citenamefont {Ponte}\ \emph
  {et~al.}(2015{\natexlab{a}})\citenamefont {Ponte}, \citenamefont {Chandran},
  \citenamefont {Papic},\ and\ \citenamefont {Abanin}}]{Ponte2015196}%
  \BibitemOpen
  \bibfield  {author} {\bibinfo {author} {\bibfnamefont {P.}~\bibnamefont
  {Ponte}}, \bibinfo {author} {\bibfnamefont {A.}~\bibnamefont {Chandran}},
  \bibinfo {author} {\bibfnamefont {Z.}~\bibnamefont {Papic}}, \ and\ \bibinfo
  {author} {\bibfnamefont {D.~A.}\ \bibnamefont {Abanin}},\ }\href {\doibase
  http://dx.doi.org/10.1016/j.aop.2014.11.008} {\bibfield  {journal} {\bibinfo
  {journal} {Annals of Physics}\ }\textbf {\bibinfo {volume} {353}},\ \bibinfo
  {pages} {196 } (\bibinfo {year} {2015}{\natexlab{a}})}\BibitemShut {NoStop}%
\bibitem [{\citenamefont {D'Alessio}\ and\ \citenamefont
  {Rigol}(2014)}]{PhysRevX.4.041048}%
  \BibitemOpen
  \bibfield  {author} {\bibinfo {author} {\bibfnamefont {L.}~\bibnamefont
  {D'Alessio}}\ and\ \bibinfo {author} {\bibfnamefont {M.}~\bibnamefont
  {Rigol}},\ }\href {\doibase 10.1103/PhysRevX.4.041048} {\bibfield  {journal}
  {\bibinfo  {journal} {Phys. Rev. X}\ }\textbf {\bibinfo {volume} {4}},\
  \bibinfo {pages} {041048} (\bibinfo {year} {2014})}\BibitemShut {NoStop}%
\bibitem [{\citenamefont {Ponte}\ \emph
  {et~al.}(2015{\natexlab{b}})\citenamefont {Ponte}, \citenamefont {Papic},
  \citenamefont {Huveneers},\ and\ \citenamefont
  {Abanin}}]{PhysRevLett.114.140401}%
  \BibitemOpen
  \bibfield  {author} {\bibinfo {author} {\bibfnamefont {P.}~\bibnamefont
  {Ponte}}, \bibinfo {author} {\bibfnamefont {Z.}~\bibnamefont {Papic}},
  \bibinfo {author} {\bibfnamefont {F.}~\bibnamefont {Huveneers}}, \ and\
  \bibinfo {author} {\bibfnamefont {D.~A.}\ \bibnamefont {Abanin}},\ }\href
  {\doibase 10.1103/PhysRevLett.114.140401} {\bibfield  {journal} {\bibinfo
  {journal} {Phys. Rev. Lett.}\ }\textbf {\bibinfo {volume} {114}},\ \bibinfo
  {pages} {140401} (\bibinfo {year} {2015}{\natexlab{b}})}\BibitemShut
  {NoStop}%
\bibitem [{\citenamefont {Lazarides}\ \emph {et~al.}(2015)\citenamefont
  {Lazarides}, \citenamefont {Das},\ and\ \citenamefont
  {Moessner}}]{LazaridesMBL}%
  \BibitemOpen
  \bibfield  {author} {\bibinfo {author} {\bibfnamefont {A.}~\bibnamefont
  {Lazarides}}, \bibinfo {author} {\bibfnamefont {A.}~\bibnamefont {Das}}, \
  and\ \bibinfo {author} {\bibfnamefont {R.}~\bibnamefont {Moessner}},\ }\href
  {\doibase 10.1103/PhysRevLett.115.030402} {\bibfield  {journal} {\bibinfo
  {journal} {Phys. Rev. Lett.}\ }\textbf {\bibinfo {volume} {115}},\ \bibinfo
  {pages} {030402} (\bibinfo {year} {2015})}\BibitemShut {NoStop}%
\bibitem [{\citenamefont {Abanin}\ \emph {et~al.}(2016)\citenamefont {Abanin},
  \citenamefont {Roeck},\ and\ \citenamefont {Huveneers}}]{AbaninDrivenMBL}%
  \BibitemOpen
  \bibfield  {author} {\bibinfo {author} {\bibfnamefont {D.~A.}\ \bibnamefont
  {Abanin}}, \bibinfo {author} {\bibfnamefont {W.~D.}\ \bibnamefont {Roeck}}, \
  and\ \bibinfo {author} {\bibfnamefont {F.}~\bibnamefont {Huveneers}},\ }\href
  {\doibase http://dx.doi.org/10.1016/j.aop.2016.03.010} {\bibfield  {journal}
  {\bibinfo  {journal} {Annals of Physics}\ }\textbf {\bibinfo {volume}
  {372}},\ \bibinfo {pages} {1 } (\bibinfo {year} {2016})}\BibitemShut
  {NoStop}%
\bibitem [{\citenamefont {Abanin}\ \emph {et~al.}(2015)\citenamefont {Abanin},
  \citenamefont {De~Roeck},\ and\ \citenamefont
  {Huveneers}}]{AbaninSlowHeating}%
  \BibitemOpen
  \bibfield  {author} {\bibinfo {author} {\bibfnamefont {D.~A.}\ \bibnamefont
  {Abanin}}, \bibinfo {author} {\bibfnamefont {W.}~\bibnamefont {De~Roeck}}, \
  and\ \bibinfo {author} {\bibfnamefont {F.~m.~c.}\ \bibnamefont {Huveneers}},\
  }\href {\doibase 10.1103/PhysRevLett.115.256803} {\bibfield  {journal}
  {\bibinfo  {journal} {Phys. Rev. Lett.}\ }\textbf {\bibinfo {volume} {115}},\
  \bibinfo {pages} {256803} (\bibinfo {year} {2015})}\BibitemShut {NoStop}%
\bibitem [{\citenamefont {{Weinberg}}\ \emph {et~al.}(2016)\citenamefont
  {{Weinberg}}, \citenamefont {{Bukov}}, \citenamefont {{D'Alessio}},
  \citenamefont {{Polkovnikov}}, \citenamefont {{Vajna}},\ and\ \citenamefont
  {{Kolodrubetz}}}]{2016arXiv160602229W}%
  \BibitemOpen
  \bibfield  {author} {\bibinfo {author} {\bibfnamefont {P.}~\bibnamefont
  {{Weinberg}}}, \bibinfo {author} {\bibfnamefont {M.}~\bibnamefont {{Bukov}}},
  \bibinfo {author} {\bibfnamefont {L.}~\bibnamefont {{D'Alessio}}}, \bibinfo
  {author} {\bibfnamefont {A.}~\bibnamefont {{Polkovnikov}}}, \bibinfo {author}
  {\bibfnamefont {S.}~\bibnamefont {{Vajna}}}, \ and\ \bibinfo {author}
  {\bibfnamefont {M.}~\bibnamefont {{Kolodrubetz}}},\ }\href@noop {} {\bibfield
   {journal} {\bibinfo  {journal} {ArXiv e-prints}\ } (\bibinfo {year}
  {2016})},\ \Eprint {http://arxiv.org/abs/1606.02229} {arXiv:1606.02229
  [cond-mat.quant-gas]} \BibitemShut {NoStop}%
\bibitem [{\citenamefont {Eckardt}\ and\ \citenamefont
  {Holthaus}(2008)}]{PhysRevLett.101.245302}%
  \BibitemOpen
  \bibfield  {author} {\bibinfo {author} {\bibfnamefont {A.}~\bibnamefont
  {Eckardt}}\ and\ \bibinfo {author} {\bibfnamefont {M.}~\bibnamefont
  {Holthaus}},\ }\href {\doibase 10.1103/PhysRevLett.101.245302} {\bibfield
  {journal} {\bibinfo  {journal} {Phys. Rev. Lett.}\ }\textbf {\bibinfo
  {volume} {101}},\ \bibinfo {pages} {245302} (\bibinfo {year}
  {2008})}\BibitemShut {NoStop}%
\bibitem [{\citenamefont {{Eckardt}}(2016)}]{2016arXiv160608041E}%
  \BibitemOpen
  \bibfield  {author} {\bibinfo {author} {\bibfnamefont {A.}~\bibnamefont
  {{Eckardt}}},\ }\href@noop {} {\bibfield  {journal} {\bibinfo  {journal}
  {ArXiv e-prints}\ } (\bibinfo {year} {2016})},\ \Eprint
  {http://arxiv.org/abs/1606.08041} {arXiv:1606.08041 [cond-mat.quant-gas]}
  \BibitemShut {NoStop}%
\bibitem [{\citenamefont {Russomanno}\ and\ \citenamefont
  {Torre}(2016)}]{0295-5075-115-3-30006}%
  \BibitemOpen
  \bibfield  {author} {\bibinfo {author} {\bibfnamefont {A.}~\bibnamefont
  {Russomanno}}\ and\ \bibinfo {author} {\bibfnamefont {E.~G.~D.}\ \bibnamefont
  {Torre}},\ }\href {http://stacks.iop.org/0295-5075/115/i=3/a=30006}
  {\bibfield  {journal} {\bibinfo  {journal} {EPL (Europhysics Letters)}\
  }\textbf {\bibinfo {volume} {115}},\ \bibinfo {pages} {30006} (\bibinfo
  {year} {2016})}\BibitemShut {NoStop}%
\bibitem [{\citenamefont {Lieb}\ and\ \citenamefont
  {Robinson}(1972)}]{Lieb1972}%
  \BibitemOpen
  \bibfield  {author} {\bibinfo {author} {\bibfnamefont {E.~H.}\ \bibnamefont
  {Lieb}}\ and\ \bibinfo {author} {\bibfnamefont {D.~W.}\ \bibnamefont
  {Robinson}},\ }\href {\doibase 10.1007/BF01645779} {\bibfield  {journal}
  {\bibinfo  {journal} {Communications in Mathematical Physics}\ }\textbf
  {\bibinfo {volume} {28}},\ \bibinfo {pages} {251} (\bibinfo {year}
  {1972})}\BibitemShut {NoStop}%
\bibitem [{\citenamefont {{Hastings}}(2010)}]{2010arXiv1008.5137H}%
  \BibitemOpen
  \bibfield  {author} {\bibinfo {author} {\bibfnamefont {M.~B.}\ \bibnamefont
  {{Hastings}}},\ }\href@noop {} {\bibfield  {journal} {\bibinfo  {journal}
  {ArXiv e-prints}\ } (\bibinfo {year} {2010})},\ \Eprint
  {http://arxiv.org/abs/1008.5137} {arXiv:1008.5137 [math-ph]} \BibitemShut
  {NoStop}%
\bibitem [{\citenamefont {Polkovnikov}(2005)}]{PhysRevB.72.161201}%
  \BibitemOpen
  \bibfield  {author} {\bibinfo {author} {\bibfnamefont {A.}~\bibnamefont
  {Polkovnikov}},\ }\href {\doibase 10.1103/PhysRevB.72.161201} {\bibfield
  {journal} {\bibinfo  {journal} {Phys. Rev. B}\ }\textbf {\bibinfo {volume}
  {72}},\ \bibinfo {pages} {161201} (\bibinfo {year} {2005})}\BibitemShut
  {NoStop}%
\bibitem [{\citenamefont {De~Grandi}\ \emph {et~al.}(2010)\citenamefont
  {De~Grandi}, \citenamefont {Gritsev},\ and\ \citenamefont
  {Polkovnikov}}]{PhysRevB.81.012303}%
  \BibitemOpen
  \bibfield  {author} {\bibinfo {author} {\bibfnamefont {C.}~\bibnamefont
  {De~Grandi}}, \bibinfo {author} {\bibfnamefont {V.}~\bibnamefont {Gritsev}},
  \ and\ \bibinfo {author} {\bibfnamefont {A.}~\bibnamefont {Polkovnikov}},\
  }\href {\doibase 10.1103/PhysRevB.81.012303} {\bibfield  {journal} {\bibinfo
  {journal} {Phys. Rev. B}\ }\textbf {\bibinfo {volume} {81}},\ \bibinfo
  {pages} {012303} (\bibinfo {year} {2010})}\BibitemShut {NoStop}%
\bibitem [{\citenamefont {{De Grandi}}\ and\ \citenamefont
  {{Polkovnikov}}(2010)}]{2010LNP...802...75D}%
  \BibitemOpen
  \bibfield  {author} {\bibinfo {author} {\bibfnamefont {C.}~\bibnamefont {{De
  Grandi}}}\ and\ \bibinfo {author} {\bibfnamefont {A.}~\bibnamefont
  {{Polkovnikov}}},\ }in\ \href {\doibase 10.1007/978-3-642-11470-0_4} {\emph
  {\bibinfo {booktitle} {Lecture Notes in Physics, Berlin Springer Verlag}}},\
  \bibinfo {series} {Lecture Notes in Physics, Berlin Springer Verlag}, Vol.\
  \bibinfo {volume} {802},\ \bibinfo {editor} {edited by\ \bibinfo {editor}
  {\bibfnamefont {A.~K.~K.}\ \bibnamefont {{Chandra}}}, \bibinfo {editor}
  {\bibfnamefont {A.}~\bibnamefont {{Das}}}, \ and\ \bibinfo {editor}
  {\bibfnamefont {B.~K.~K.}\ \bibnamefont {{Chakrabarti}}}}\ (\bibinfo {year}
  {2010})\ p.~\bibinfo {pages} {75},\ \Eprint {http://arxiv.org/abs/0910.2236}
  {arXiv:0910.2236 [cond-mat.stat-mech]} \BibitemShut {NoStop}%
\bibitem [{\citenamefont {Hone}\ \emph {et~al.}(1997)\citenamefont {Hone},
  \citenamefont {Ketzmerick},\ and\ \citenamefont {Kohn}}]{Kohn97}%
  \BibitemOpen
  \bibfield  {author} {\bibinfo {author} {\bibfnamefont {D.~W.}\ \bibnamefont
  {Hone}}, \bibinfo {author} {\bibfnamefont {R.}~\bibnamefont {Ketzmerick}}, \
  and\ \bibinfo {author} {\bibfnamefont {W.}~\bibnamefont {Kohn}},\ }\href
  {\doibase 10.1103/PhysRevA.56.4045} {\bibfield  {journal} {\bibinfo
  {journal} {Phys. Rev. A}\ }\textbf {\bibinfo {volume} {56}},\ \bibinfo
  {pages} {4045} (\bibinfo {year} {1997})}\BibitemShut {NoStop}%
\bibitem [{foo()}]{footnote1}%
  \BibitemOpen
  \href@noop {} {}\bibinfo {note} {We thank Max Metlitski for providing this
  argument.}\BibitemShut {Stop}%
\end{thebibliography}%


\begin{thebibliography}{16}

\expandafter\ifx\csname natexlab\endcsname\relax\def\natexlab#1{#1}\fi
\expandafter\ifx\csname bibnamefont\endcsname\relax
  \def\bibnamefont#1{#1}\fi
\expandafter\ifx\csname bibfnamefont\endcsname\relax
  \def\bibfnamefont#1{#1}\fi
\expandafter\ifx\csname citenamefont\endcsname\relax
  \def\citenamefont#1{#1}\fi
\expandafter\ifx\csname url\endcsname\relax
  \def\url#1{\texttt{#1}}\fi
\expandafter\ifx\csname urlprefix\endcsname\relax\def\urlprefix{URL }\fi
\providecommand{\bibinfo}[2]{#2}
\providecommand{\eprint}[2][]{\url{#2}}


\bibitem{Santos15}



\bibitem{deutsch} J. M. Deutsch, Phys. Rev. A {\bf 43}, 2046 (1991).
\bibitem{srednicki} M. Srednicki, Phys. Rev. E {\bf 50}, 888 (1994).
\bibitem{Rigol08} M. Rigol, V. Dunjko, and M. Olshanii, Nature {\bf 452},854 (2008). 




\bibitem{polkovnikovreview}
M. Bukov, L. D'Alessio and A.Polkovnikov,  Advances in Physics, Vol. 64, No. 2, 139-226 (2015) 


%
%
%
%
%
%


%
%
%
%
%
%
%






\end{thebibliography}

\appendix 
\section{Bounds on $||W^{(k)}||$ and $||\Omega^{(k)}(t)||$}
We derive bounds on the norms of $||W^{(k)}||$ and $||\Omega^{(k)}(t)||$. We start with a generalization of Eqn.~(\ref{eqn:Omega}),
\begin{align}
& \Omega^{(k)}(t) = - i \int_{-\infty}^t dt' e^{pst'}( V^{(k)}(t') - W^{(k)} ), \text{ and } \nonumber \\
& W^{(k)} = p s \int_{-\infty}^0 dt' e^{p s t'} V^{(k)}(t').
\end{align}
We assume that $V^{(k)}(t)$ is periodic, not necessarily with zero time average, and we would like to show that $|| W^{(k)} || \sim ||V^{(k)}||$, and also $|| \Omega^{(k)}(t) || \sim \frac{1}{\omega}||V^{(k)}||$, up to multiplicative corrections of powers of $ s/\omega $.  We have
\begin{align}
|| W^{(k)} || &= \left|\left| ps \int_{-\infty}^0 dt' e^{pst'} V^{(k)}(t') \right|\right| \nonumber \\
& =\left|\left|  \bar{V}^{(k)} + \frac{ps}{1-e^{-psT}}  \int_{-T}^0 dt' e^{pst'} \tilde{V}^{(k)}(t')  \right|\right| \nonumber \\
& = \left|\left|  \bar{V}^{(k)} -\frac{ps}{1-e^{-psT}} \int_{-T}^0 dt ps e^{pst} \int_0^t dt' \tilde{V}^{(k)}(t')  \right|\right| \nonumber \\
& \leq \left(||\bar{V}^{(k)}|| +  ||\tilde{V}^{(k)}||\frac{ps}{1-e^{-psT}}\int_{-T}^0 dt ps e^{pst} T \right) \nonumber \\
& =  ||\bar{V}^{(k)}|| + ||\tilde{V}^{(k)}||(psT)   \nonumber \\
& \leq 
\begin{cases}
||V^{(k)} ||\left(\frac{ps}{\omega} \right)  \qquad \text{ if } \bar{V}^{(k)} = 0 \\
||V^{(k)} ||\left(1 + \frac{ps}{\omega} \right) ~\text{ if } \bar{V}^{(k)} \neq 0
\end{cases},
\end{align}
where $\bar{V} = \frac{1}{T}\int_0^T V(t) dt$ and $\tilde{V}(t) = V(t) - \bar{V}$. In the second line, we have used the fact that for a time-periodic term, one can split the integral into an infinite series $\int_{-T}^0 + \int^{-T}_{-2T} + \int^{-2T}_{-3T} + \cdots$, shift the argument of the integrand, and sum the resulting geometric series. In the third line, we have used an integration by parts, and in the fourth line, the triangle inequality and the inequality $|| \int f  || \leq \int ||f||$. Next,
\begin{align}
& || \Omega^{(k)} (t) || =  \left| \left| \int_{-\infty}^t dt' e^{pst'} (V^{(k)}(t') - W^{(k)} ) \right| \right| \nonumber \\
& = e^{pst} \left| \left| \int_{-\infty}^0 dt' e^{pst'} \left( \tilde{V}^{(k)} (t'+t) - ( W^{(k)} - \bar{V}^{(k)})  \right) \right| \right| \nonumber \\
& \leq e^{pst}  \left( \left| \left| \int_{-\infty}^0 dt' e^{pst'} \tilde{V}^{(k)}(t'+t) \right|\right| + \frac{1}{ps} || W^{(k)} - \bar{V}^{(k)} ||  \right) \nonumber \\
& \leq 2 e^{pst} || \tilde{V}^{(k)}|| \left( \frac{1}{\omega} \right) \leq 2 e^{pst} || V^{(k)}|| \left( \frac{1}{\omega} \right),
\end{align}
using the previous result regarding $W^{(k)}$. Since $e^{pst} \leq 1$ during the ramp, we have
\begin{align}
|| W^{(k)} || \sim ||V^{(k)}||, \qquad || \Omega^{(k)}(t) || \sim \frac{1}{\omega}||V^{(k)}||,
\end{align}
up to multiplicative corrections of $s/\omega$.

\section{Scaling of difference of measurements of a local operator $\delta \langle O \rangle$}
In this appendix we derive the scaling of the difference of two measurements of a local observable $O$, one by a state $\rho$ evolved by $H'(t)$, and one by the same state $\rho$ but evolved with $H_{\rm eff}^{(s)}(t)$. We call this difference $\delta \langle O \rangle$,
\be
\delta \langle O \rangle  \equiv \left|\text{Tr}\left(U(0) \rho U^\dagger(0) O\right) - \text{Tr}\left(U_{\rm eff}^{(s)} (0) \rho U_{\rm eff}^{(s) \dagger}(0) O\right) \right|,
\ee
where  $U(t)$ and $U_{\rm eff}^{(s)}(t)$ are time evolution operators satisfying
\begin{align}
& i \frac{d }{dt} U(t) = H'(t) U(t), U(-\infty) = \mathbb{I}, \nonumber \\
&i \frac{d }{dt} U_{\rm eff}^{(s)} (t) = H_{\rm eff}^{(s)}(t) U_{\rm eff}^{(s)}(t), U_{\rm eff}^{(s)}(-\infty) = \mathbb{I}.
\end{align}

We begin by looking at the object
\begin{align}
f(t) \equiv U_{\rm eff}^{(s) \dagger}(t) U(t) \rho U^\dagger(t) U_{\rm eff}^{(s)}(t).
\end{align}
Now,
\begin{align}
\frac{d}{dt} \left( U_{\rm eff}^{(s)}(t) U(t) \right) = -i U_{\rm eff}^{(s)} (t) H_{\rm fast}(t) U(t) .
\end{align}
We take the time derivative of $f(t)$ and then integrate from $t' = -\infty$ to $0$:
\begin{align}
&f(0) - f(-\infty)  = \int_{-\infty}^{0} dt' \frac{df(t')}{dt'},
\end{align}
which gives 
\begin{align}
& U_{\rm eff}^{(s) \dagger}(0) U(0) \rho U^\dagger(0) U_{\rm eff}^{(s)}(0) - \rho  \nonumber \\
& =  - i \int_{-\infty}^0 dt' U_{\rm eff}^{(s) \dagger}(t') [H_{\rm fast}(t'), U(t') \rho U^\dagger(t') ] U_{\rm eff}^{(s) }(t'). 
\end{align}
Multiplying the above expression on the left by $U_{\rm eff}^{(s)}(0)$ and on the right by $U_{\rm eff}^{(s) \dagger }(0)$ gives us $U(0) \rho U^\dagger (0)  - U_{\rm eff}^{(s)} (0) \rho U_{\rm eff}^{(s) \dagger }(0)$, i.e. the difference between the final density matrices at the end of the evolution.

Then,
\begin{widetext}
\begin{align}
\delta \langle O \rangle & \equiv \left|\text{Tr}\left(U(0) \rho U^\dagger(0) O\right) - \text{Tr}\left(U_{\rm eff}^{(s)} (0) \rho U_{\rm eff}^{(s) \dagger}(0) O\right) \right| \nonumber \\
& = \left| \int_{-\infty}^0 dt' \text{Tr} \left( U_{\rm eff}^{(s)}(0) U_{\rm eff}^{(s) \dagger}(t') [H_{\rm fast}(t'), U(t') \rho U^\dagger(t') ] U_{\rm eff}^{(s) }(t') U_{\rm eff}^{(s) \dagger}(0) O    \right) \right| \nonumber \\
& = \left| \int_{-\infty}^0 dt' \text{Tr} \left( U(t') \rho U^\dagger(t') [ U_{\rm eff}^{(s) }(t') U_{\rm eff}^{(s) \dagger}(0) O U_{\rm eff}^{(s) \dagger } U_{\rm eff}^{(s)}(t') ,H_{\rm fast}(t')]  \right)\right| \nonumber \\
& \leq \int_{-\infty}^0 dt' \left|\left| [ U_{\rm eff}^{(s) }(t') U_{\rm eff}^{(s) \dagger}(0) O U_{\rm eff}^{(s) \dagger } U_{\rm eff}^{(s)}(t') ,H_{\rm fast}(t')]  \right|\right| \nonumber  \\
& =   \int_{-\infty}^0 dt' \left| \left| [\mathcal{U}^{(s)}_{\rm eff}(t') O \mathcal{U}^{(s) \dagger}_{\rm eff}(t'), H_{\rm fast}(t')] \right| \right| ,
\end{align}
\end{widetext}
where in the third line we have used the cyclicity of the trace to move operators around, and $\mathcal{U}^{(s)}_{\rm eff}(t) \equiv U_{\rm eff}^{(s) }(t) U_{\rm eff}^{(s) \dagger}(0)$. This object satisfies the differential equation
\begin{align}
i \frac{d}{dt} \mathcal{U}^{(s)}_{\rm eff}(t) = H_{\rm eff}^{(s)}(t) \mathcal{U}^{(s)}_{\rm eff}(t), \mathcal{U}^{(s)}_{\rm eff}(0) = \mathbb{I},
\end{align}
so it is a time evolution operator from $t = 0$ to $t$. Since $H_{\rm eff}^{(s)}(t)$ is a local many-body Hamiltonian, with typical local energy scale $J$, it has a Lieb-Robinson velocity $v_{LR} \sim J$, and therefore the object $\mathcal{U}^{(s)}_{\rm eff}(t) O \mathcal{U}^{(s) \dagger}_{\rm eff}(t)$ grows in physical extent within its lightcone $x \sim v_{LR} t$. That is, if $O$ is initially supported in a spatial region $X$, then after time $t$ it would have physically spread to become an operator supported in a ball of radius $v_{LR} t$ centered around $X$. 

Now, since $H_{\rm fast}(t')$ is a quasi-local Hamiltonian (that is, it is a sum of terms with larger and larger support $|X|$, but which has a factor that decreases at least exponentially fast $\sim e^{-\alpha |X|}$, its behavior can be extracted from the largest term in $H_{\rm fast}(t')$, which is $H^{(n_*)}(t')$, which has a range $R n_*$ with amplitude $\sim e^{-r n_*} \times e^{s t'}$ (see Refs.~\cite{2015arXiv151003405A, 2015arXiv150905386A}).  Since $n_* \sim \omega$ (Eq.~(\ref{eqn:OptimalN})), we then have
\begin{align}
\delta \langle O \rangle &\leq \sim  e^{- r n_*} \int_{-\infty}^0 dt' e^{st'} (v_{LR} |t'| + R n_* ) \nonumber \\
& \sim e^{-\mathcal{C} \frac{\omega}{J}}  s^{-2},
\end{align}
where we have kept only the scaling behavior of this quantity with $s$ and $\omega$. More carefully treating the quasi-local nature of $H_{\rm fast}(t')$ would involve modifying the constant $\mathcal{C}$ to one that depends on the decay constant associated with the quasi-local nature of the Hamiltonian, see Ref.~\cite{2015arXiv150905386A}.

\section{Determining the dominant relevant operator at the QCP}
We consider the slow quench protocol from $t = -\infty$ to $0$,
\be
H_0 + \lambda(t) \mathcal{O}_1 + \lambda(t)^2 \mathcal{O}_2,
\label{eqn:QCP}
\ee
where $H_0$ is sitting at a QCP, and $\mathcal{O}_1$, $\mathcal{O}_2$ are relevant operators pulling the system away from criticality. We assume that they have dissimilar scaling dimensions $\Delta_{1}$ and $\Delta_{2}$, that are being turned on at different rates $\lambda(t)$ and $\lambda(t)^2$ respectively, with $\lambda(t) = e^{st}$. Note that $\mathcal{O}_2$ is being ramped up at a slower rate than $\mathcal{O}_1$. 

In general, the behavior of the system near the QCP, such as  the energy scale, the correlation length, etc., will be a scaling function of critical exponents associated with both relevant operators. However, because of the difference in ramp speeds of $\mathcal{O}_1$ and $\mathcal{O}_2$, such quantities will generically instead be dominated by just one of the two relevant operators.

For example, in the case when $\mathcal{O}_1$ is a more relevant operator than $\mathcal{O}_2$, then since $\lambda(t) > \lambda(t)^2$ for the duration of the ramp, it is clear that the scaling behavior of $\mathcal{O}_1$ dictates the energy and length scales in the problem. However, in the case when $\mathcal{O}_2$ is the more relevant operator, the situation is not as clear : $\mathcal{O}_2$, while being more relevant, carries with it a smaller prefactor of $\lambda^2(t)$, which might or might not make it more dominant than $\lambda(t) \mathcal{O}_1$.

Here we determine precisely which operator, $\mathcal{O}_1$ or $\mathcal{O}_2$ is the dominant relevant operator that governs the scaling functions of the energy scale, correlation length, etc.~near the QCP in Eqn.~(\ref{eqn:QCP}). We argue as follows \cite{footnote1}. Let us call $\lambda_{\mathcal{O}_1} \equiv \lambda$ and $\lambda_{\mathcal{O}_2} \equiv \lambda^2$; then, their respective scaling dimensions are $d+1  - \Delta_1$ and $d+1 - \Delta_2$ respectively, where $d$ is the spatial dimension. We then need to compare the ratio
\be
\lambda_{\mathcal{O}_1}^{\frac{1}{d+1 - \Delta_1}} / \lambda_{\mathcal{O}_2}^{\frac{1}{d+1 - \Delta_2}} = \lambda^\alpha,
\ee
which compares two objects associated with $\mathcal{O}_1$ and $\mathcal{O}_2$ that each has scaling dimension $1$. Here the exponent
\begin{align}
\alpha \equiv \frac{1}{d+1 - \Delta_1} -\frac{2}{d+1 - \Delta_2} .
\end{align}
 Now, the conditions are clear: if $\lambda^\alpha > 1$, then $\lambda \mathcal{O}_1$ governs the behavior of the system near the QCP, while if $\lambda^\alpha < 1$, then $\lambda^2 \mathcal{O}_2$ governs the behavior of the system near the QCP. After simplifying, we have the resulting criteria: 
\begin{align}
\Delta_1 < \frac{d+1}{2} + \frac{\Delta_2}{2}
\end{align}
implies $\mathcal{O}_1$  is dominant while
\be
\Delta_1 > \frac{d+1}{2} + \frac{\Delta_2}{2},
\ee
implies $\mathcal{O}_2$ is dominant. Notice that the `clear' scenario discussed, when the QCP's behavior should be governed by $\mathcal{O}_1$ if it is the more relevant operator than $\mathcal{O}_2$ to begin with,  i.e.~$\Delta_1 < \Delta_2$, automatically satisfies the inequality above.

\end{document}